\documentclass{article}
\usepackage[english]{babel}
\usepackage[utf8]{inputenc}
\usepackage{johd}
\usepackage{ragged2e}
\usepackage{amsmath}
\usepackage{array}
\usepackage{graphicx}
\usepackage{caption}
\usepackage{booktabs}
\usepackage{amssymb}
\usepackage{pifont}
\usepackage{natbib}
\usepackage{rotating}
\usepackage[export]{adjustbox}
\usepackage{wrapfig}
\title{Advancing Portfolio Optimization: Adaptive Minimum-Variance Portfolios and Minimum Risk Rate Frameworks\footnote{We extend our gratitude to the Department of Mathematics and Statistics at Texas Tech University for providing access to the Bloomberg Terminal, which facilitated the extraction of the data utilized in this study.}}

\author{Ayush Jha$^{1}$$^{*}$, Abootaleb Shirvani$^{2}$, Ali Jaffri$^{1},$ Svetlozar T. Rachev$^{3}$ and Frank J. Fabozzi$^{4}$ \\
        \small $^{1}$Department of Economics, Texas Tech University \\
        \small $^{2}$Department of Mathematical Sciences, Kean University \\
        \small $^{3}$Department of Mathematics and Statistics, Texas Tech University \\
        \small $^{4}$Carey Business School, Johns Hopkins University \\\\
        \small $^{*}$Corresponding author: Ayush Jha; \tt{ayush.jha@ttu.edu} \\
}
\date{}

\begin{document}
\maketitle
\begin{abstract} 
\noindent This study presents the Adaptive Minimum-Variance Portfolio (AMVP) framework and the Adaptive Minimum-Risk Rate (AMRR) metric, innovative tools designed to optimize portfolios dynamically in volatile and nonstationary financial markets. Unlike traditional minimum-variance approaches, the AMVP framework incorporates real-time adaptability through advanced econometric models, including ARFIMA-FIGARCH processes and non-Gaussian innovations. Empirical applications on cryptocurrency and equity markets demonstrate the proposed framework's superior performance in risk reduction and portfolio stability, particularly during periods of structural market breaks and heightened volatility. The findings highlight the practical implications of using the AMVP and AMRR methodologies to address modern investment challenges, offering actionable insights for portfolio managers navigating uncertain and rapidly changing market conditions.
  \end{abstract}

\noindent\\


\section{Introduction} \label{Intro}
\justifying
\setlength{\parskip}{10pt}

Safe assets play several essential roles in financial markets, including enabling institutions to meet regulatory requirements, serving as primary pricing benchmarks, providing collateral in financial transactions, and forming a cornerstone in the theory of pricing assets and derivatives, which relies on the assumption of a safe or riskless asset. However, the expansion of financial assets has outpaced the supply of traditional safe assets. This imbalance is largely driven by non-bank financial institutions, commonly referred to as ``shadow institutions,” which operate under fewer regulatory constraints than depository banks. Unlike depository banks, shadow institutions are not required to maintain equivalent financial reserves relative to their market exposures, allowing them to take on higher levels of leverage. This structural vulnerability became alarmingly evident during the 2008 Global Financial Crisis.

Evidence from \citet{Gorton2012} demonstrates that while the overall percentage of safe debt in the U.S. economy has remained stable at approximately 32\%, the composition of these assets has shifted significantly. Traditional instruments like government debt and demand deposits have given way to innovative financial products such as money market mutual funds, commercial paper, repurchase agreements, and securitized debt, including asset-backed and mortgage-backed securities. This shift reflects the increasing role of the shadow banking system in creating synthetic safe assets. The evolving nature of these assets highlights a key limitation in financial markets: most investor portfolios, consisting of $N$ risky assets, operate without a conventional risk-free asset as a hedging instrument (see \citet{Black1972}, \citet{hansenjag}, and more recently \citet{Caballero2021}). This absence leads to market incompleteness, where investors cannot fully hedge against systemic and idiosyncratic risks.

To address the absence of conventional safe assets, \citet{Rachev2017} proposed a theoretical construct for deriving a shadow riskless rate and its corresponding riskless asset, even in markets composed entirely of risky assets. This approach enables the construction of complete and arbitrage-free systems. This riskless rate, referred to as the shadow riskless rate, is derived as the (negative of) drift term of the state-price deflator for the market, ensuring that the resulting system of risky assets and the perpetual derivative becomes complete and arbitrage-free. This innovation provides a foundation for considering portfolio optimization in incomplete markets by dynamically adapting to nonlinear dependencies and non-Gaussian characteristics in asset returns. By addressing these challenges, the shadow riskless rate motivates a shift toward optimization method, such as the Adaptive Minimum Variance Portfolio, that identify minimum-risk configurations while accommodating the evolving dynamics of a market.

Building on the critical roles of safe assets and the challenges posed by their absence in portfolios—particularly in the derivation of a riskless rate—the present paper introduces the Adaptive Minimum Variance Portfolio method. In this method, synthetic assets are iteratively generated based on the expected return and volatility of a pool of risky assets and are incorporated into the portfolio until it achieves a specified threshold variance. This iterative process enables a systematic reduction of the variance, resulting in portfolios with minimal variance rather than relying solely on classical mean-variance optimization. The Adaptive Minimum Variance Portfolio method provides a robust alternative to traditional methods by dynamically adapting to the nonlinear dependencies and non-Gaussian characteristics commonly observed in financial data. We validate this approach using two distinct classes of assets: the 30 stocks in the Dow Jones Industrial Average and the 30 largest (in terms of market capitalization) cryptocurrencies. Through this method, we derive a static measure from the iterative Adaptive Minimum Variance Portfolio process, referred to as the Adaptive Minimum Risk Rate, which serves as a shadow riskless rate, addressing the absence of a conventional risk-free asset in the portfolio.

Using the Adaptive Minimum Variance Portfolio method, we also construct a time series of Adaptive Minimum-Risk Rates, dynamically recalibrated over rolling windows to capture the evolution of risk--return trade-offs in the portfolio over time. This time series incorporates changes in asset-specific volatilities, correlations, and expected returns, allowing it to reflect varying market conditions such as periods of heightened uncertainty, shifts in liquidity, and structural breaks in the financial system. Unlike traditional risk-free rates, which are static and often influenced by policy-driven factors, the Adaptive Minimum-Risk Rate derived through the Adaptive Minimum Variance Portfolio is market-driven and capable of adapting to the real-time dynamics in the underlying asset universe. By capturing these variations, the Adaptive Minimum-Risk Rate not only provides a forward-looking measure of the safety of a portfolio but also serves as a benchmark for assessing systemic risk and evaluating alternative investment strategies.

The Adaptive Minimum-Risk Rate method offers a more econometrically robust measure of the shadow riskless rate compared to more recent approaches focusing on Principal Component Analysis (PCA)-based methods, such as those  in \citet{ADRIAN2013110}, \citet{wuxia}, and \citet{LauriaEtAl2024}. Unlike PCA, which relies on a fixed covariance structure and is sensitive to eigenvalue instability in high-dimensional settings, the Adaptive Minimum-Risk Rate is dynamically recalculated with each introduction of a synthetic asset, allowing it to adapt to evolving market conditions. This dynamic recalibration directly incorporates nonlinear dependencies and real-world constraints, such as no short-selling, making it versatile for diverse asset classes, including those with complex volatility structures like cryptocurrencies.

The shadow risk-free rate is referred to as a ``shadow” because it emerges endogenously within a portfolio of $N+1$ assets, where one asset is a synthetic, minimum-variance asset derived as a linear combination of the other $N$ assets. This synthetic asset is indivisible from the portfolio due to residual terms inherent in the linear combination, which serve as sources of additional diversification. These residuals enable the Adaptive Minimum Variance Portfolio to achieve a lower minimum variance than traditional methods, particularly in markets with extreme tail risks and nonlinear relations. Consequently, the Adaptive Minimum-Risk Rate reflects a more granular optimization and provides a dynamic, market-driven benchmark that is both theoretically sound and practical for modern financial modeling.

The indivisibility of the synthetic asset arises because its construction inherently accounts for the portfolio’s overall risk structure, including the residual terms that capture unhedged risks and nonlinear dependencies between the $N$ assets. These residuals, far from being negligible, act as channels of additional diversification, effectively lowering the portfolio’s minimum achievable variance. This is particularly significant in markets characterized by non-Gaussian features such as heavy tails and skewness, where traditional methods struggle to take into account the complex interdependencies. By incorporating these residual effects, the Adaptive Minimum Variance Portfolio method ensures that the Adaptive Minimum-Risk Rate represents a risk-free benchmark that is both internally consistent with the portfolio’s dynamics and reflective of market realities.

The development of a shadow riskless rate, or Adaptive Minimum Risk Rate, derived from portfolios of assets in the Dow Jones Industrial Average and cryptocurrencies with lower long-range dependence (LRD) than traditional riskless rates, such as the 10-year Treasury Bond rate and 3-month Treasury Bond rate, is a novel contribution in the literature. This approach addresses the limitations of traditional government bonds by constructing a riskless rate dynamically from diversified portfolios of equities and cryptocurrencies. The inclusion of cryptocurrencies, which are highly volatile, have extremely non-Gaussian distributions, and relatively short market histories, is particularly novel and unprecedented. No previous work has systematically incorporated such diverse classes of assets to derive a proxy for riskless rates, let alone demonstrated the capacity of this approach to achieve lower LRD. Traditional riskless rates exhibit near-unit-root persistence (e.g., $d(v) = 0.9999$), which limits their responsiveness to changing market conditions. In contrast, this paper, by deriving the Adaptive Minimum-Risk Rate, constructed using adaptive portfolio methods and simulated scenarios such as Normal Inverse Gaussian (NIG)-distributed innovations (serving as the benchmark for generating forward-looking data), achieves a significantly lower LRD. For the first time in the literature, we find a minimum-risk rate that achieves lower persistence in volatility and higher expected returns than the traditional ``risk-free" rate used widely in macroeconomic and financial models. This forward-looking, market-driven measure enables more robust applications to portfolio optimization, macroeconomic modeling, and asset pricing.

The Adaptive Minimum-Risk Rate offers a dynamic and market-driven measure of portfolio risk, paving the way for novel applications, including the construction of a volatility index derived from portfolios of cryptocurrencies. Such an index would reflect the realized volatility influenced by the nonlinear dependencies and extreme market behaviors inherent in these classes of assets,  mitigating limitations of traditional measures of dispersion, such as the VIX. While the proposed index may not trade one-to-one against the VIX, it could complement traditional volatility measures, providing unique insights into markets experiencing uncertainty shocks, particularly when these shocks extend beyond the scope of conventional equity indices and the class of digital assets (see \citet{shirvani}, \citet{jha}).

Moreover, the Adaptive Minimum-Risk Rate's ability to achieve higher expected returns and lower persistence in volatility than traditional riskless rates highlights its potential to be a more adaptive benchmark for hedging strategies. In portfolios of \(N\) risky assets, it provides a forward-looking and dynamically adjusted risk-free proxy that better aligns with modern financial environments, where cryptocurrencies and other nontraditional assets are playing a growing role. The present paper offers a theoretically consistent and empirically grounded proxy for addressing the absence of a true riskless rate while aligning with modern asset pricing models that rely on short- to medium-term predictability. The explicit reduction in the LRD, coupled with the ability to dynamically handle non-traditional classes of assets, represents an area largely unexplored in the existing literature, positioning this research as a cornerstone for future advances in asset pricing, portfolio management, and financial modeling.

The structure of this paper is as follows: Section \ref{Lit Review} provides a comprehensive review of the relevant literature, with a focus on the expanding body of work surrounding shadow riskless rates and the specific gaps this study seeks to address. Section \ref{Methods} details the methodological framework employed for constructing the Adaptive Minimum Variance Portfolio, Adaptive Minimum Risk Rate, and Adaptive Minimum-CVaR Portfolio, as well as the approach used to identify structural breaks in the time series of minimum-risk rates. Section \ref{results} presents the empirical findings, including estimates of long-range dependence and the unique characteristics of the Adaptive Minimum-Risk Rate compared to traditional riskless rates. Finally, Section \ref{conclusion} summarizes the key contributions of this study and outlines potential avenues for future research.

\section{Relevant Work} \label{Lit Review}

This study contributes to the expanding literature on the construction of a shadow riskless rate and its implications for macroeconomic and financial modeling. \citet{Black1972} advanced financial theory by demonstrating that the Capital Asset Pricing Model (CAPM) remains valid even without a riskless asset—a framework commonly known as Black’s zero-beta CAPM. However, subsequent studies have primarily assumed the existence of a riskless asset in equilibrium models, as shown in \citet{Nielson1990}, \citet{Allingham1991}, \citet{KonnoShirakawa1995}, and \citet{SunYang2003}. These models often rely on the single-period mean-variance framework for capital asset pricing under uncertainty, as developed by \citet{sharpe}, \citet{Lintner}, and \citet{Mossin}. Nevertheless, other works have explored capital asset pricing structures under the more realistic assumption that all assets are risky.

In this context, \citet{Lintner69}, \citet{Fama}, and \citet{Black1972} adopt the pragmatic view that a single, universal interest rate at which investors and firms can borrow and lend unlimited amounts does not exist in practice. Empirical evidence supports this perspective, showing that consumers typically borrow at higher interest rates than they lend, while firms often secure loans at more favorable rates than individual consumers. These observations highlight the presence of risk premia in loan markets, particularly for loans of equivalent maturity, under assumptions of market efficiency and consumer rationality.

Building on this theoretical foundation, \citet{mayers} introduced a framework for capital asset pricing in the absence of a riskless asset, accounting for the role of nonmarketable assets. This model differs from traditional CAPM frameworks in two important ways. First, it explicitly incorporates risks attributable to nonmarketable assets into measures of systematic risk for both firms and the market portfolio. Second, it removes the unrealistic assumption that all maximizing investors hold identical portfolios. Even with homogeneous expectations, the model predicts that investors construct distinct portfolios of marketable assets, offering a more realistic depiction of investment behavior.

Further advancing these ideas, \citet{hansenjag} derived restrictions that security market data impose on intertemporal marginal rates of substitution (IMRS) in dynamic economic models. Their framework establishes bounds on the mean and standard deviation of the IMRS using asset prices, extending analysis to a broader class of models while avoiding restrictive assumptions about model parameters. The shadow riskless rate, constructed from the expected IMRS in a no-arbitrage environment, emerges as a theoretical construct in portfolios without explicit risk-free instruments. This IMRS-based approach aligns closely with the Adaptive Minimum Risk Rate (AMRR) method, particularly in accommodating non-Gaussian characteristics.

Similarly, \citet{ACHARYA2005375} developed a liquidity-adjusted CAPM (L-CAPM) that models the absence of a single risk-free asset, such as government bonds, through the interplay of liquidity and return dynamics. Their framework reveals that liquidity premia play a critical role in imperfect markets and that shadow rates can dynamically reflect changing market conditions. By capturing the persistence and predictability of illiquidity, the L-CAPM offers a foundation for using shadow rates as tools for hedging against liquidity-driven volatility, especially for portfolios containing high-transaction-cost or illiquid assets.

Turning to the demand for U.S. Treasury bonds, \citet{kvj} analyzes their dual attributes of liquidity and safety, demonstrating that Treasuries serve as substitutes for money by offering liquidity and security comparable to cash. The high demand for these bonds generates seigniorage-like benefits for the government by lowering its borrowing costs. However, in markets where the availability of true risk-free assets is limited, constructing shadow risk-free rates becomes essential. The convenience yield framework outlined by \citet{kvj} highlights the premium investors pay for liquidity and safety—attributes that must be reflected in shadow rates. Furthermore, the dynamic pricing of Treasuries, driven by supply conditions, aligns closely with the Adaptive Minimum-Risk Rate (AMRR) method proposed in this paper, which dynamically incorporates evolving market conditions into its construction.

\citet{duffie} provide a complementary perspective through their reduced-form framework for valuing defaultable securities, focusing on the term structure of yields for corporate and sovereign bonds. Their model adjusts the short-rate process to account for expected losses due to default by incorporating a hazard rate process. In markets where truly risk-free assets are unavailable, their framework demonstrates how defaultable bonds can act as substitutes, with adjusted short-rate processes reflecting implicit shadow risk-free rates that incorporate default risks. This dynamic adjustment of short rates mirrors the method used in constructing shadow rates in this study, where portfolio weights and covariance structures adapt to changing market conditions.

\citet{NBERw7346} extend the affine pricing framework traditionally applied to bond term structures to equity pricing, creating a unified model that prices both stocks and bonds using consistent state variables. These variables capture fundamental economic uncertainties, such as inflation, dividend growth, and risk aversion, enabling the derivation of a shadow risk-free rate that adapts to macroeconomic dynamics. In contrast, \citet{ADRIAN2013110} proposes a regression-based approach to modeling term structures, emphasizing computational efficiency and flexibility. Their method constructs risk-free benchmarks by isolating term structure dynamics and linking bond pricing to broader economic indicators, providing an effective substitute for riskless rates in environments where such assets are absent.

Together, these contributions establish the theoretical and empirical foundations for constructing shadow risk-free rates. By incorporating elements such as liquidity, default risk, and macroeconomic dynamics, these studies support the development of robust methodologies adapted to non-Gaussian market environments and the limited availability of truly risk-free assets, closely aligning with the approaches introduced in this paper.

\section{Methods} \label{Methods}
\subsection{Data} \label{Data}

We construct two distinct portfolios for our analysis: one comprising the 30 constituents of the Dow Jones Industrial Average (DJIA) and the other consisting of 30 major cryptocurrencies (by market capitalization). All data were retrieved from Bloomberg Terminal on 20 November 2024, and the DJIA composition reflects the constituents as of that date. The equity portfolio, focusing on DJIA components, spans from 23 November 2020 through 18 November 2024. This window provides approximately four years of daily data, with the U.S. equity market operating on roughly 252 trading days per year. In contrast, the cryptocurrency portfolio extends from 19 November 2020 through 17 November 2024, also comprising about four years of daily observations\footnote{Summary statistics of the Portfolio of Assets in DJIA and Cryptocurrencies is presented in Appendix \ref{B}}. Because cryptocurrencies trade continuously (including weekends and holidays), we approximate a frequency of 365 trading days per year.

We use unadjusted closing prices for equities as reported by Bloomberg prior to the ex-dividend date at the end of the fourth quarter, meaning our price series do not incorporate the downward price adjustments that typically follow dividend payouts. Consequently, dividend-related effects are not reflected in the return series. For cryptocurrencies, we record one daily closing price per asset from Bloomberg, thus maintaining consistency across the sample. We compute arithmetic returns for both asset classes, as arithmetic returns enable straightforward aggregation of portfolio returns using linear weights and align well with mean-variance or minimum-variance optimization models.

On the historical, or backward-looking, front, we perform a minimum-variance optimization directly on the historical returns of each asset in both portfolios, relying on the assumption that historical covariances and volatilities provide an adequate basis for future risk estimation. However, to capture potential future market dynamics beyond what the historical data alone might reveal, we fit an Autoregressive Fractionally Integrated Moving Average–Fractionally Integrated Generalized Autoregressive Conditional Heteroskedasticity model, ARFIMA(1, \(d(m)\), 1)-FIGARCH(1, \(d(v)\), 1), which captures long-memory effects in both the mean and conditional volatility processes. The empirical specification of the ARFIMA-FIGARCH model is in appendix \ref{ARFIMA}.

Using the parameter estimates from ARFIMA-FIGARCH, we generate an equal number of simulated returns as the historical sample size. To account for heavy tails and skewness commonly observed in both equity and cryptocurrency markets, we transform the ARFIMA-FIGARCH innovations via a NIG distribution. This approach enables modeling of rare but significant market movements and thus provides a richer set of return outcomes than standard Gaussian assumptions. By incorporating these NIG scenarios, we produce a forward-looking distribution of returns that reflects both short-term fluctuations---through the FIGARCH volatility process---and long-range dependence in mean---through ARFIMA. We then use this expanded scenario set to perform minimum-variance optimization and stress testing, thereby enhancing the robustness of the portfolio construction process. Hence, the combination of historical data and simulation-based scenarios offers a more comprehensive perspective on the risk-return profile for both equity and cryptocurrency portfolios, ultimately improving the robustness of our portfolio optimization and risk management strategies.

\subsection{Adaptive Minimum-Variance Portfolio} \label{AMVP}

We begin by considering the classical Minimum Variance Portfolio (MVP) for \(N\) risky assets observed over \(T\) trading days. We follow the method outlined in \citet{Markowitz1952} and \citet{Markowitz1959} for portfolio optimization using variance as the risk measure, therefore, adding it to identify the portfolio with minimum risk. Let \(\boldsymbol{\Sigma}_t\) denote the sample covariance matrix of asset returns at iteration \(t\), and let \(\mathbf{w}_t\) be the associated portfolio weights. The MVP optimization is explained in appendix \ref{AAMVP}.

\noindent By obtaining \(\mathbf{w}_t\) and the portfolio variance \(\sigma_t^2\), we identify the unique global MVP, i.e., the portfolio on the Markowitz efficient frontier that attains the lowest possible level of risk subject to the constraint that there is a fully invested budget and that the weights are non-negative. Unlike general mean-variance portfolios, which trade off expected returns and variance, the MVP optimizes only the variance and does not incorporate expected returns into the objective function. Consequently, its variance is strictly less than or equal to the variance of any other fully invested portfolio in the feasible set, thus guaranteeing minimal risk.

In the unconstrained setting (i.e., with possible short sales), the closed-form solution, derived in Eq. (2) in appendix \ref{AAMVP}, holds whenever the covariance matrix \(\boldsymbol{\Sigma}_t\) is invertible. However, the additional non-negativity constraint (\(\mathbf{w}_t \geq 0\)) can render the answer given by the closed-form expression infeasible if it produces negative weights. In such cases, solving the quadratic program numerically is required. Nevertheless, when non-negativity is satisfied, the resulting portfolio has the lowest possible volatility among all feasible portfolios. This property makes the MVP particularly robust to estimation errors in expected returns since its construction does not depend on them and relies solely on the second moments (covariances). Moreover, it is a corner solution on the lower boundary of the mean-variance frontier and can serve as a building block for constructing other efficient portfolios, including those with specific risk preferences or target returns.

\begin{figure}[h]
    \centering
    \begin{minipage}[t]{0.49\textwidth} 
        \centering
        \includegraphics[width=\textwidth]{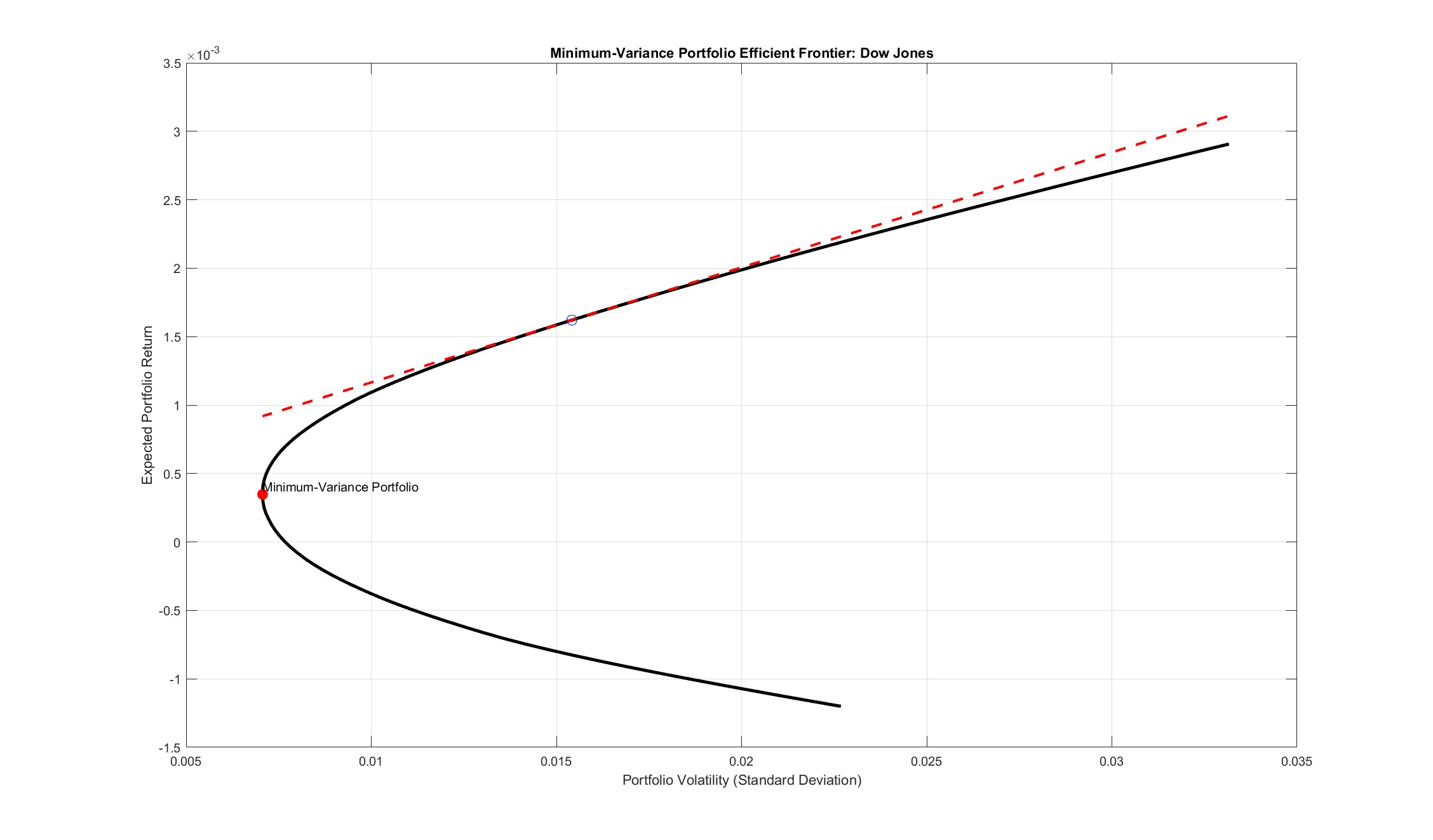} 
        \label{fig:mvp_dow}
    \end{minipage}
    \hfill
    \begin{minipage}[t]{0.49\textwidth} 
        \centering
        \includegraphics[width=\textwidth]{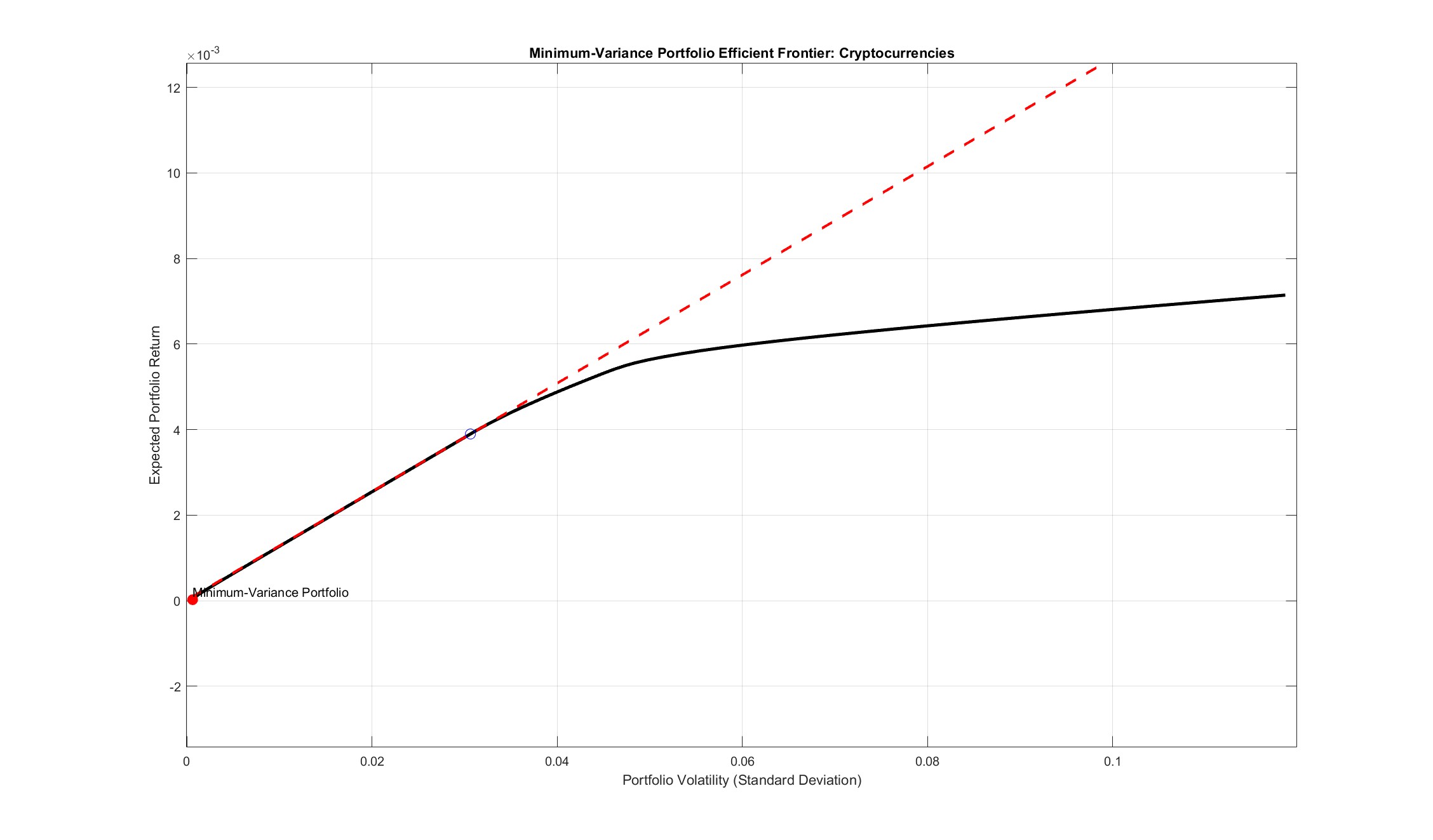} 
        \label{fig:mvp_c}
    \end{minipage}
    \caption{Minimum-Variance Portfolio based on the portfolio of Dow Jones stocks and Cryptocurrencies}
    \label{fig:MVP}
\end{figure}

Figure \ref{fig:MVP} plots the two efficient frontiers obtained from the minimum-variance portfolio optimization, each depicting the trade-off between expected return ($y$-axis) and portfolio volatility ($x$-axis), for different asset universes. The first plot shows the efficient frontier for the 30 Dow Jones stocks, while the second plot shows the same for the 30 largest (in terms of market capitalization) cryptocurrencies. In each plot, the solid black curve represents the continuum of feasible portfolios under varying weight allocations, with the left-most point on this curve marking the MVP.

Since there is no conventional risk-free asset in this analysis, the red dashed line in each figure serves as an \textit{alternate} Capital Market Line (CML), originating from the MVP rather than from a conventional risk-free rate of return. Drawing the CML in this manner helps illustrate how, in the absence of a truly risk-free instrument, one might view the trade-off between expected returns and volatility if the MVP is treated as the baseline ``zero-risk” anchor. We use the time series of the newly constructed Adaptive Minimum-Risk Rate, or the shadow risk-free rate, to build the alternate CML.\footnote{Section \ref{AMRR - main} outlines the method of constructing the Adaptive Minimum-Risk Rate; the construction of the alternate CML will be further discussed in the following section.} As we move upward along the black curve (or along the dashed line, if leveraging is allowed), higher expected returns become attainable only at the expense of higher portfolio volatility. This core relationship—observed in both equity and cryptocurrency markets—highlights the fundamental principle of mean-variance efficiency: the portfolio offering the absolute minimum risk is uniquely determined (the MVP), while all other efficient portfolios lie above that point on the risk–return trade-off curve.

\begin{figure}[h]
    \centering
    \begin{minipage}[t]{0.49\textwidth} 
        \centering
        \includegraphics[width=\textwidth]{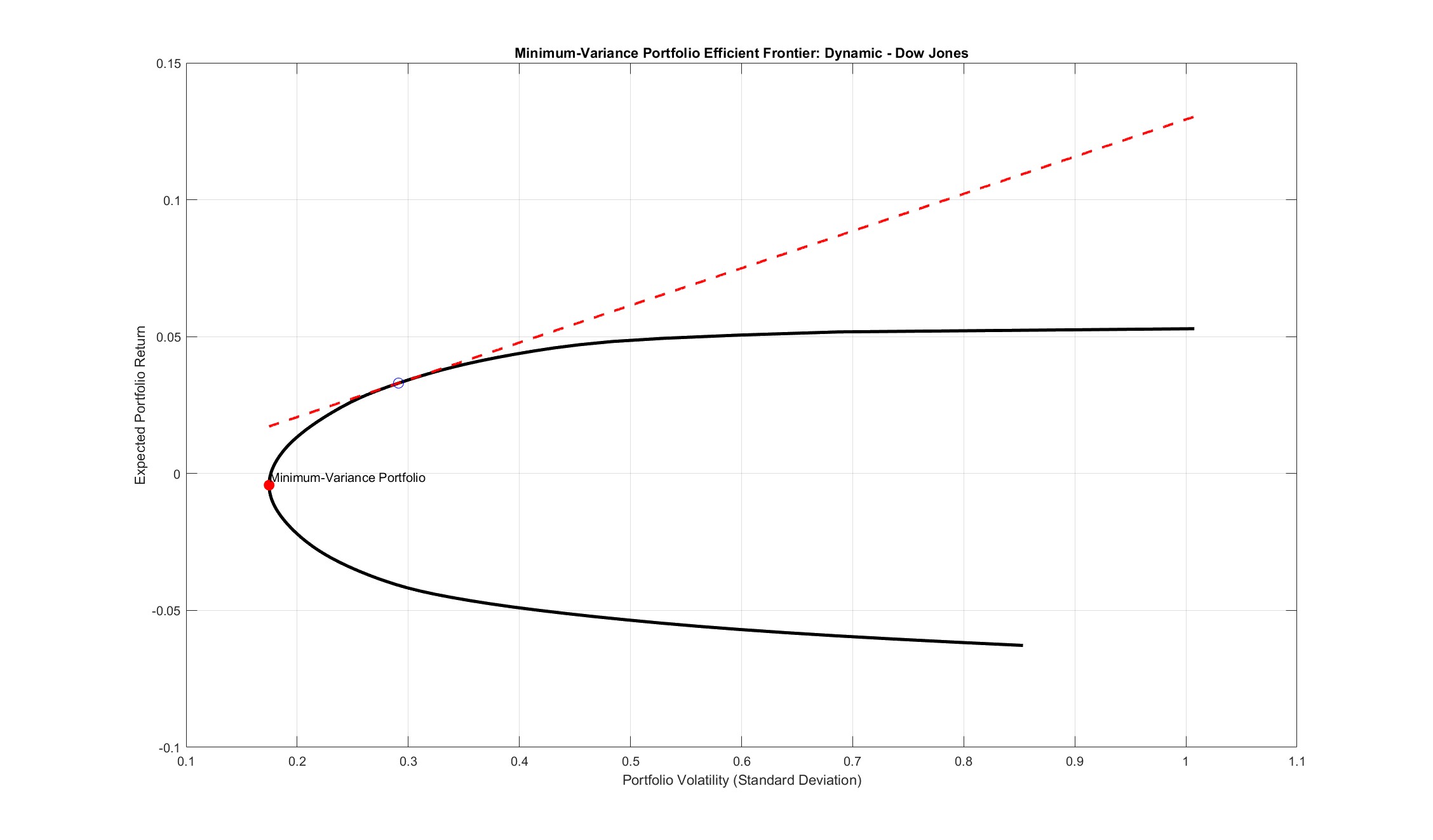} 
        \label{fig:mvp_ddow}
    \end{minipage}
    \hfill
    \begin{minipage}[t]{0.49\textwidth} 
        \centering
        \includegraphics[width=\textwidth]{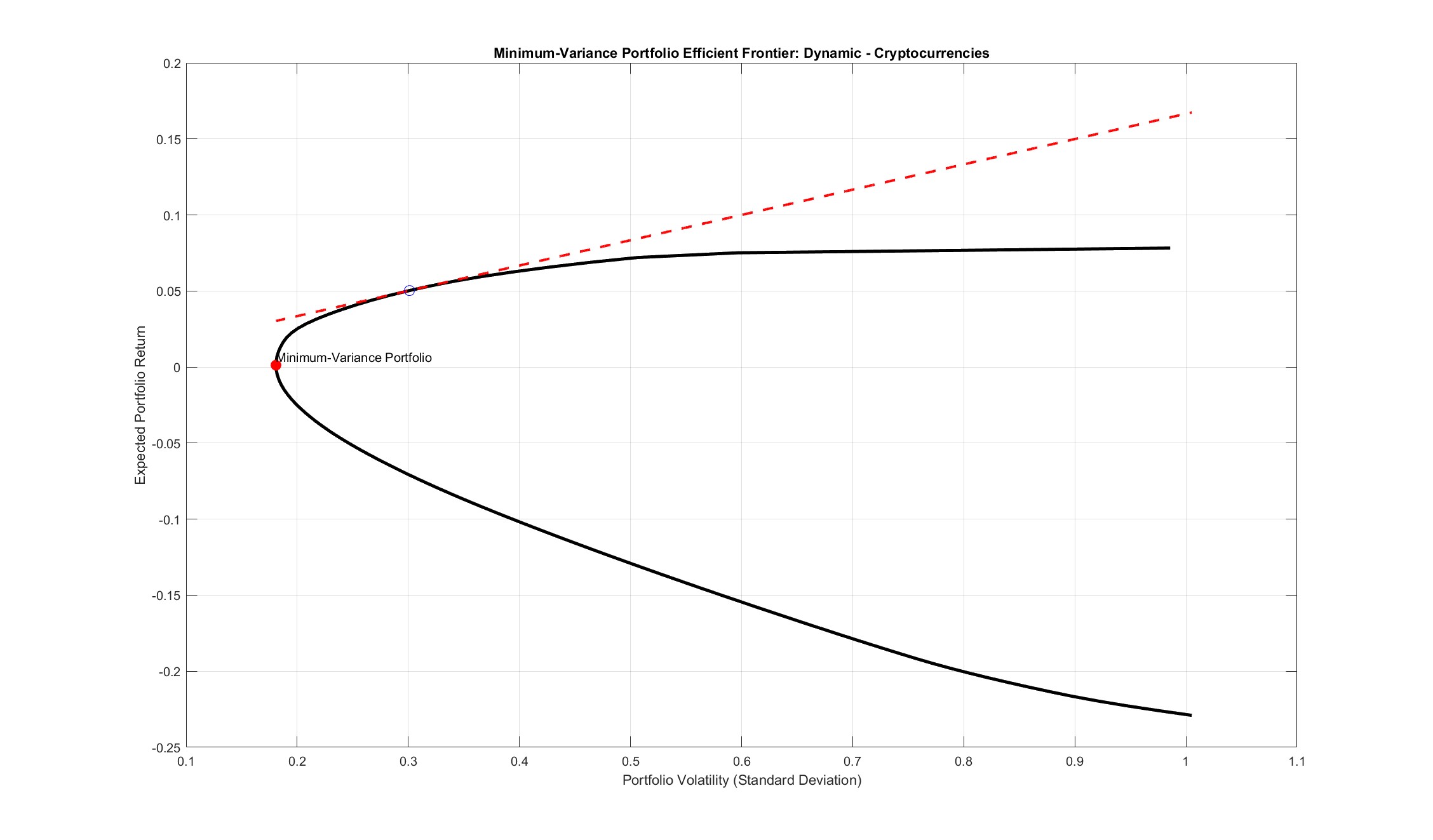} 
        \label{fig:mvp_dc}
    \end{minipage}
    \caption{Minimum-Variance Portfolio based on the \emph{forward-looking} portfolio of Dow Jones stocks and Cryptocurrencies}
    \label{fig:MVP Dynamic}
\end{figure}

Figure \ref{fig:MVP Dynamic} plots the efficient frontiers generated under the ``dynamic” approach, meaning that the returns are not drawn purely from historical observations but rather NIG scenarios generated using the normal innovations of the  ARFIMA-FIGARCH model (as described in Section \ref{Data}). By modeling the same number of observations as in the historical dataset, these ``dynamic” frontiers capture both long-memory behavior and the heavy tails often seen in equity and cryptocurrency markets. The left panel shows the Dow Jones frontier, while the right panel depicts the same for the 30 largest cryptocurrencies. In each figure, the solid black curve represents the feasible set of risk–return combinations derived from the NIG-based scenarios, with the MVP located at the left-most point on the curve.

Moreover, the red dashed line acts as an \textit{alternate} CML that originates at the MVP. This visually demonstrates the same logic of the ``zero-risk” anchor, where various combinations of return and risk can be attained through hypothetical ``leveraging” along the red dashed line. As with the historical frontiers, higher expected returns come at the cost of higher volatility, though here the simulation-based approach better accommodates the possibility of extreme market events and longer-range dependencies in asset returns.

\noindent Given the method for constructing the MVP and the efficient frontiers for both portfolios, we extend the procedure to build the adaptive minimum-variance portfolio. We initialize our procedure with a dataset \(\mathbf{R}_1 \in \mathbb{R}^{T \times N}\), containing \(T\) daily returns for \(N\) assets. Based on \(\mathbf{R}_1\), we compute the sample covariance \(\boldsymbol{\Sigma}_1\) and obtain the initial MVP weights. To refine the variance further, we introduce an iterative scheme that expands the investable universe with synthetic assets derived from each MVP solution. The method used to expand the portfolio by incorporating synthetic assets into it and updating the covariance matrix is explained in appendix \ref{AAMVP}.

\begin{figure}[htbp]
    \centering
    \begin{minipage}[t]{0.49\textwidth} 
        \centering
        \includegraphics[width=\textwidth]{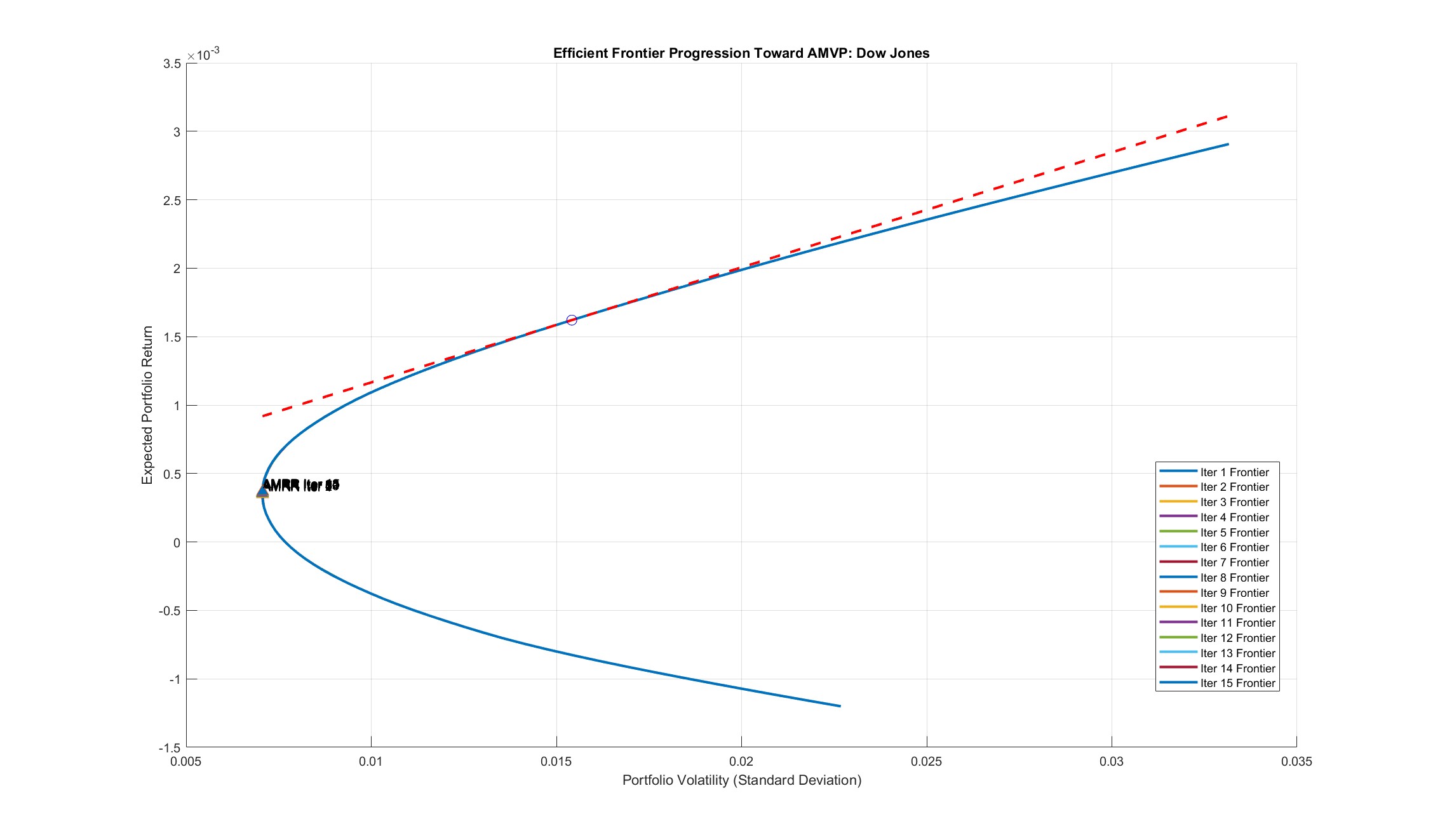} 
        \label{fig:amvp_d}
    \end{minipage}
    \hfill
    \begin{minipage}[t]{0.49\textwidth} 
        \centering
        \includegraphics[width=\textwidth]{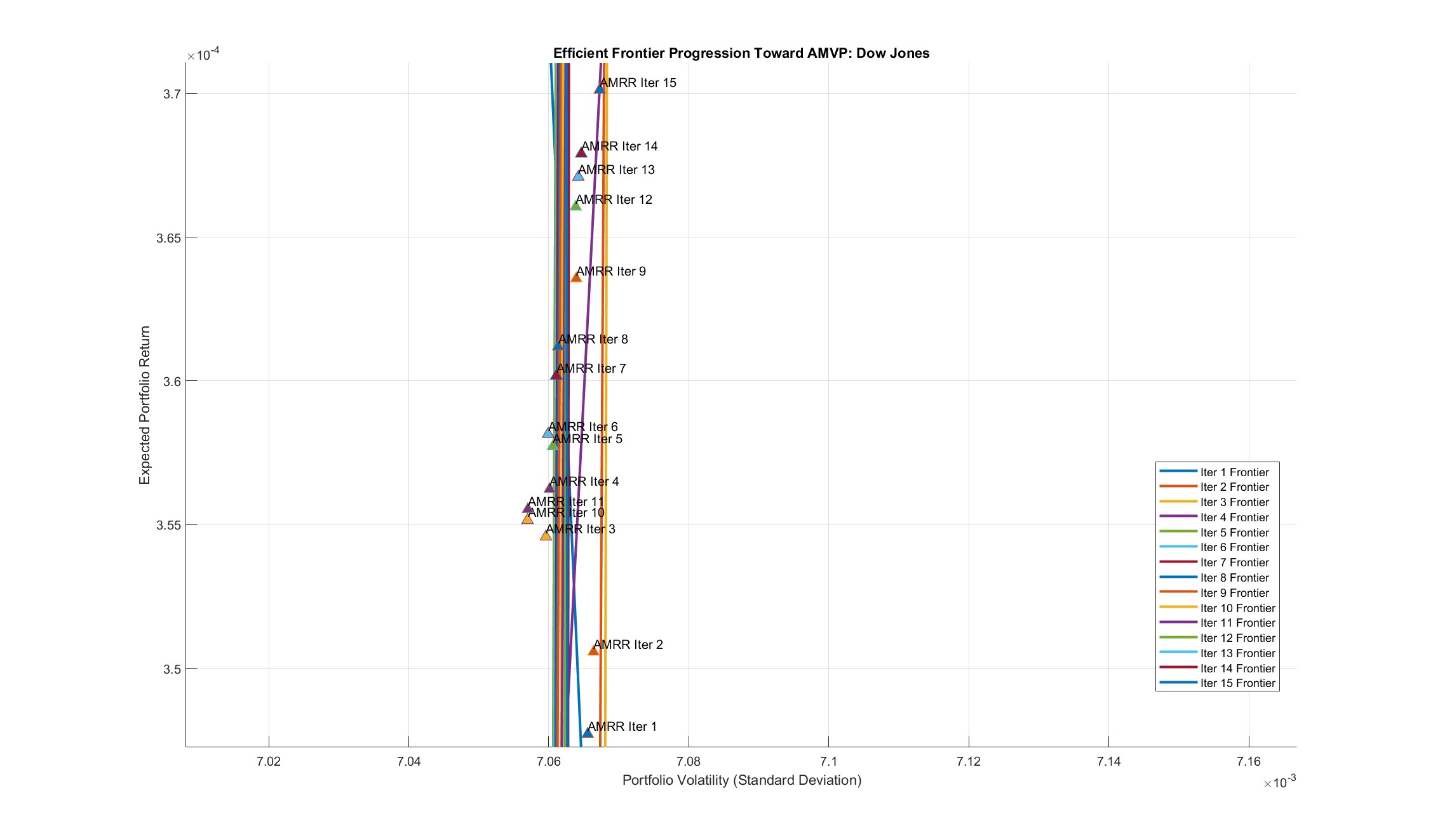} 
        \label{fig:amvp_dz}
    \end{minipage}
    \caption{Adaptive Minimum-Variance Portfolio based on the portfolio of Dow Jones stocks}
    \label{fig:amvp dow}
\end{figure}

The AMVP method provides a dynamic, iterative approach to portfolio optimization that mitigates several limitations inherent in traditional static methods. It begins by computing the mean returns and covariance matrix of an initial dataset of asset returns, such as those from the Dow Jones 30 stocks. Using these estimates, the method identifies the initial MVP, forming the baseline configuration. A key innovation of the AMVP method lies in its creation and inclusion of synthetic assets during subsequent iterations. Specifically, the MVP's weighted returns at iteration \(k\) are treated as a new synthetic asset to be added to the asset universe in iteration \(k+1\). This iterative expansion of the dataset incorporates the benefits of the results of prior optimizations into the subsequent steps, enabling a continuous refinement of the portfolio weights.

In each iteration, an efficient frontier is constructed, depicting the trade-off between portfolio variance (or standard deviation) and expected return for different weight allocations. From this frontier, the iteration's MVP is determined and its expected return recorded as the Adaptive Minimum Risk Rate. The process continues until the improvement in the variance from one iteration to the next is negligible, indicating convergence. The final AMVP is a portfolio configuration that achieves the lowest possible variance given the available assets and synthetic assets accumulated over the iterations.

Figure \ref{fig:amvp dow} illustrates the iterative progression of the AMVP over 15 iterations using the portfolio assets constituting the DJIA. In the first figure, each iteration shifts the efficient frontier inward, demonstrating reduced variance for a given level of expected return as the portfolio converges to the final AMVP. The Adaptive Minimum-Risk Rate at each step serves as a dynamic benchmark for the minimum-variance portfolio. In the second figure, a more granular view shows the convergence over the last few iterations, highlighting the diminishing gains as the algorithm nears an optimal risk configuration.

By continually adapting through the inclusion of synthetic assets, the AMVP differs from static methods and accounts more comprehensively for interdependencies between assets. This iterative mechanism is particularly robust to non-stationary market conditions, as it progressively adjusts the composition of the portfolio in response to the evolving structures of the returns and correlations. The Adaptive Minimum-Risk Rate, updated at each iteration, functions as a time-varying benchmark that offers a forward-looking measure of the portfolio’s risk profile.

\begin{figure}[htbp]
    \centering
    \begin{minipage}[t]{0.49\textwidth} 
        \centering
        \includegraphics[width=\textwidth]{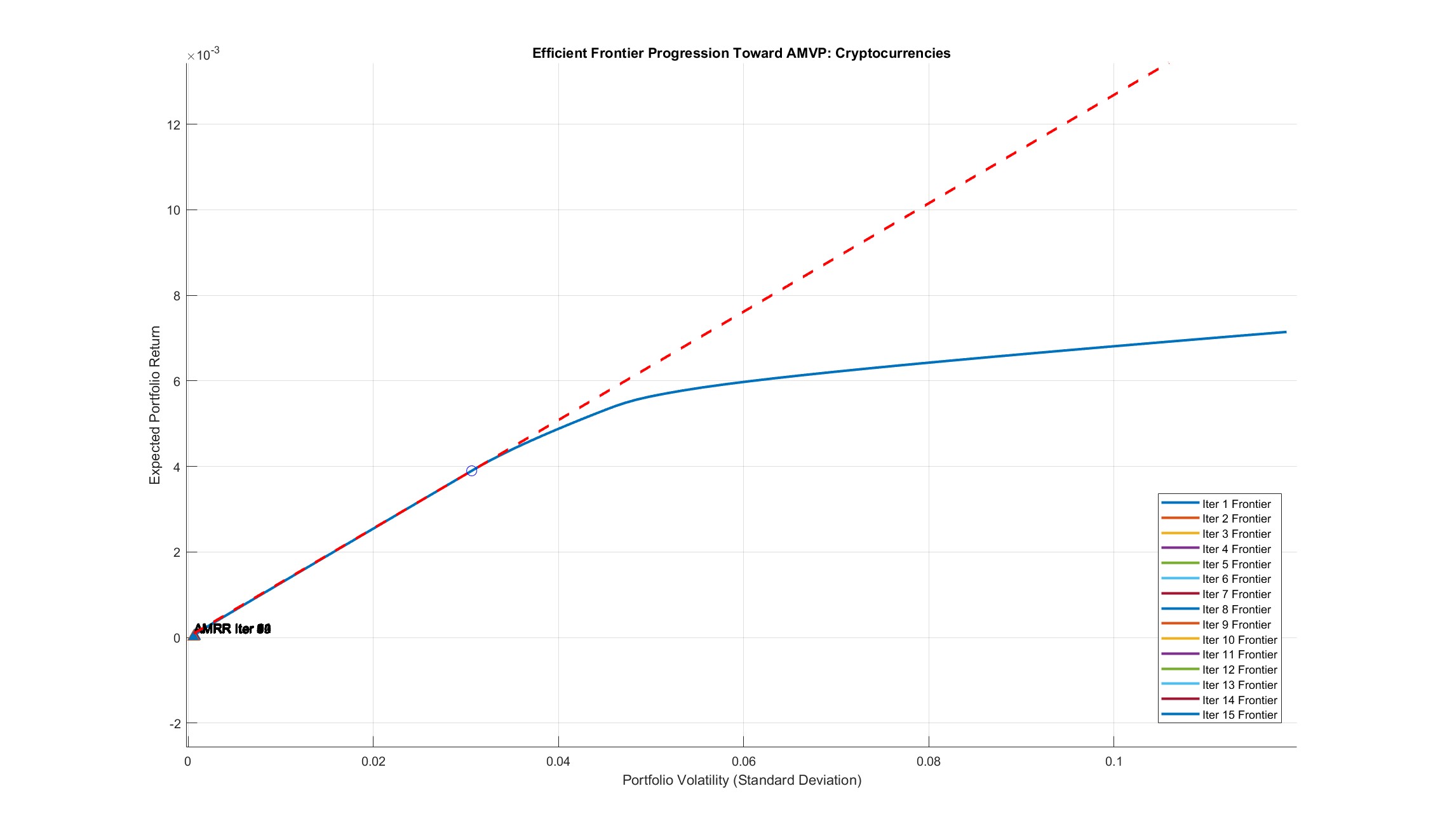} 
        \label{fig:amvp_c}
    \end{minipage}
    \hfill
    \begin{minipage}[t]{0.49\textwidth} 
        \centering
        \includegraphics[width=\textwidth]{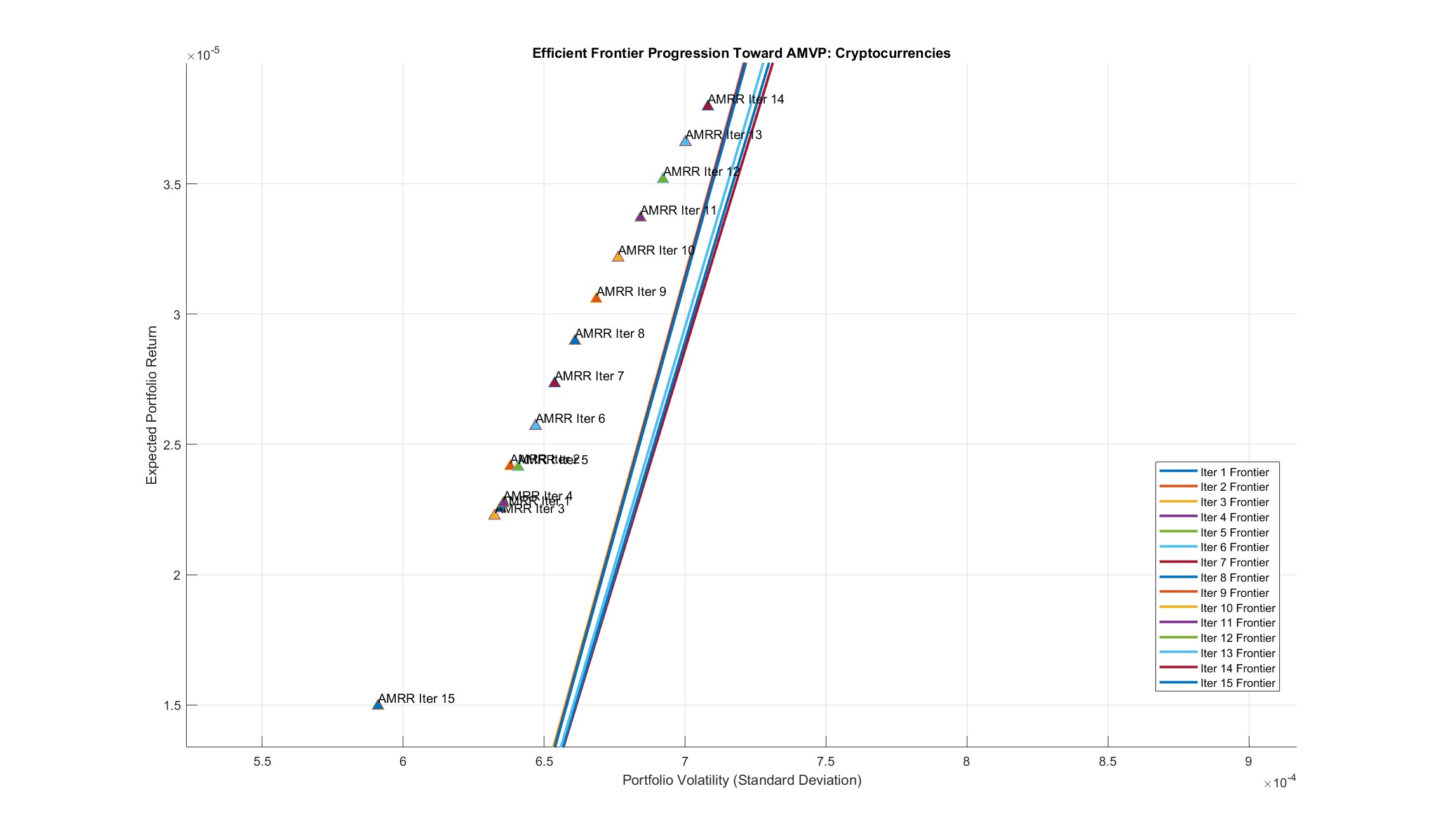} 
        \label{fig:amvp_cz}
    \end{minipage}
    \caption{Adaptive Minimum-Variance Portfolio based on the portfolio of Cryptocurrencies}
    \label{fig:amvp crypto}
\end{figure}

Figure \ref{fig:amvp crypto} plots the progression of the AMVP for a universe of cryptocurrencies, analogous to the Dow Jones examples but tailored to digital assets. In the first plot, each colored line is the efficient frontier at a given iteration. As the iteration count increases, the incorporation of synthetic assets (generated by the AMVP procedure) gradually shifts the frontier, reflecting a systematic reduction in portfolio variance for a given expected return. The red dashed line again serves as an alternate CML, originating at the AMVP in the absence of a conventional risk-free asset.

The second plot takes a closer look at the same iterative process, focusing on the progression of the Adaptive Minimum-Risk Rate as each new minimum variance portfolio emerges. Here, each triangle corresponds to the Adaptive Minimum-Risk Rate from a particular iteration, plotted together with its respective frontier. The clustering of lines and points at the higher iterations illustrates the diminishing gains in variance reduction, thereby evidencing convergence to the final AMVP. As with the case of equities, these figures highlight the robustness of the iterative optimization process, revealing how the AMVP method  dynamically refines the composition of the portfolio to achieve continually lower variance under evolving market conditions.

\begin{figure}[htbp]
    \centering
    \begin{minipage}[t]{0.49\textwidth} 
        \centering
        \includegraphics[width=\textwidth]{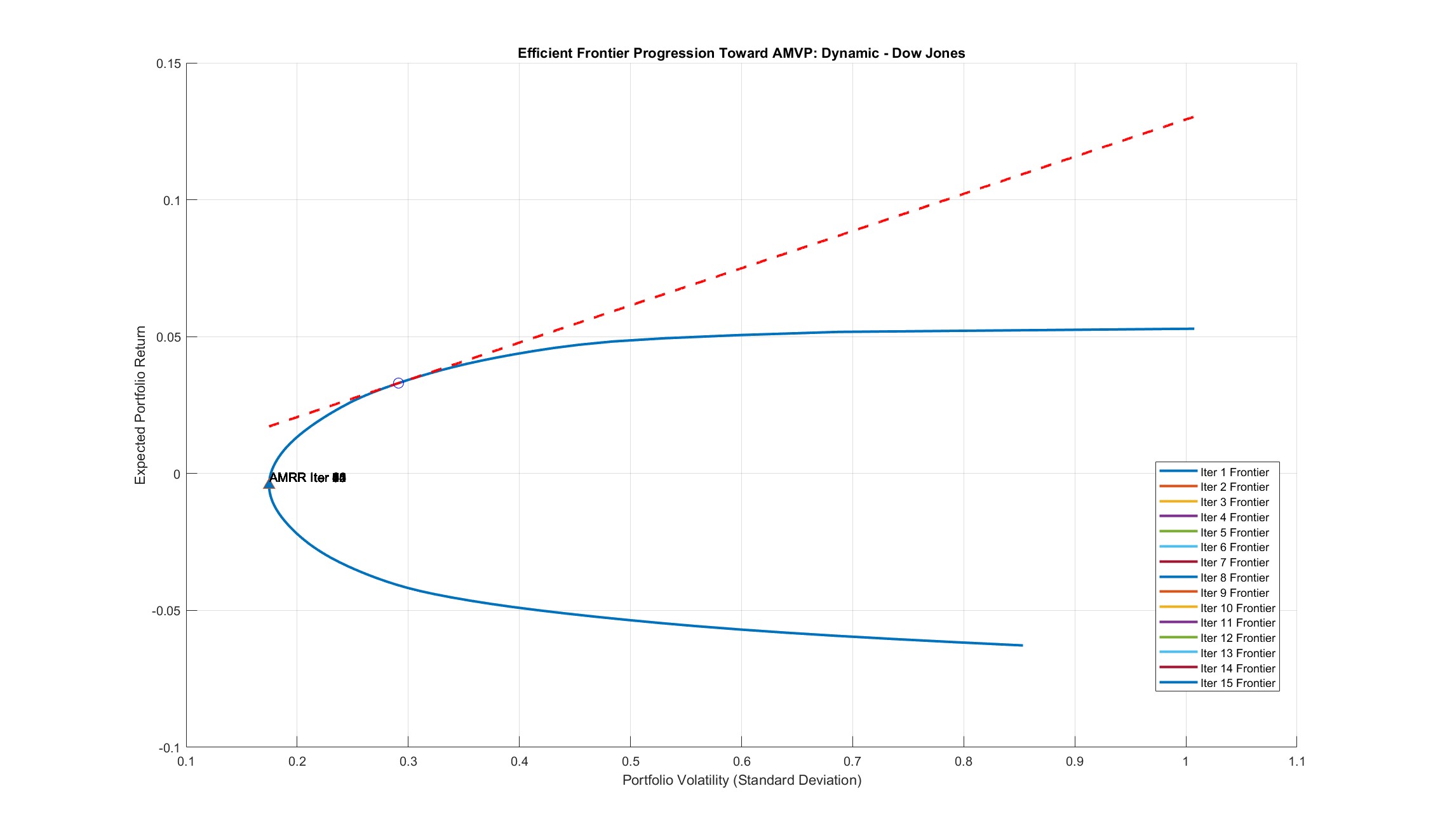} 
        \label{fig:amvp_dd}
    \end{minipage}
    \hfill
    \begin{minipage}[t]{0.49\textwidth} 
        \centering
        \includegraphics[width=\textwidth]{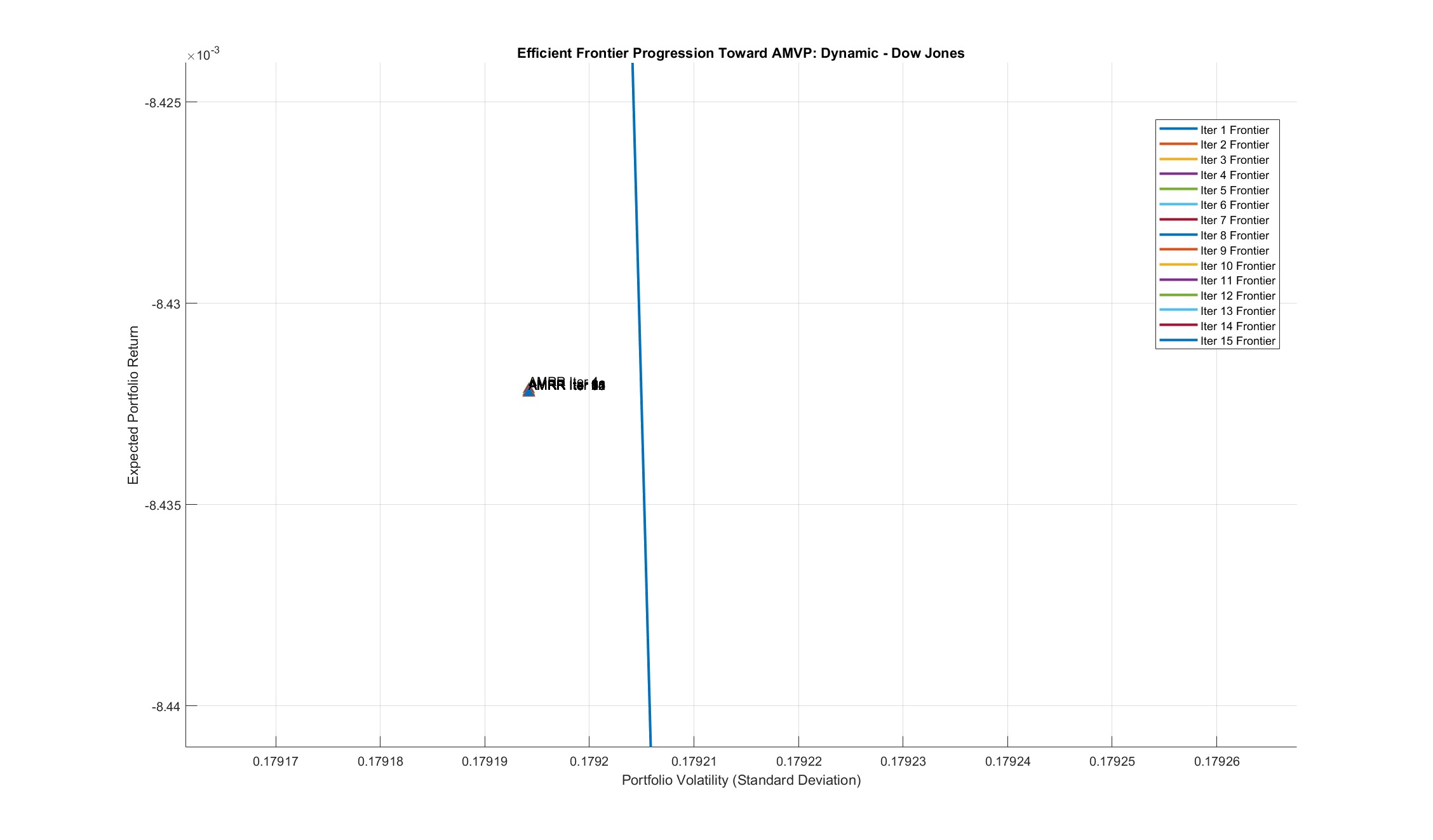} 
        \label{fig:amvp_ddz}
    \end{minipage}
    \caption{Adaptive Minimum-Variance Portfolio based on the \emph{forward-looking} portfolio of Dow Jones stocks}
    \label{fig:amvp dyn dow}
\end{figure}

Figure \ref{fig:amvp dyn dow} illustrates the progression of the AMVP for the Dow Jones under a dynamic, forward-looking regime. In contrast to the historically based efficient frontiers, these frontiers reflect scenarios generated by fitting an ARFIMA-FIGARCH model and then simulating normal innovations via an NIG distribution. This allows the optimization to account for heavy tails, skewness, and long-memory effects that may not be fully captured in historical data. In the first plot, each colored line represents the efficient frontier at a given iteration. As synthetic assets (derived from the MVP’s weighted returns in the preceding iteration) are iteratively added to the asset universe, the frontier shifts, reflecting increasingly refined diversification opportunities under the simulated forward-looking conditions. The red dashed line again denotes an alternate CML, derived from the respective portfolio's Adaptive Minimum-Risk Rate, given that no conventional risk-free asset is assumed.

The second plot provides a more granular view, zooming in on the final iterations to highlight the convergence of the Adaptive Minimum-Risk Rate. Each marker corresponds to the rate obtained at one of the later iterations, while the blue line traces the associated frontier. Notably, this dynamic approach produces an efficient frontier that may differ in both shape and location from its historically based counterpart—chiefly due to the inclusion of potential extreme events and evolving correlations. The eventual convergence to a single MVP confirms that the AMVP procedure remains robust even under simulated market conditions with heavier tails and non-stationary dynamics, offering a forward-looking perspective on the construction of a minimum-risk portfolio.

\begin{table}[ht]
    \centering
    \begin{tabular}{cccc}
    \hline
    \hline
        \textbf{Shadow Rate} & \textbf{Annualized} & \textbf{\textit{Synthetic Assets}} & \textbf{Observations}\\
        \hline
        \hline
        DJIA AMVP & 0.093 & 10 & 752\\
        Cryptocurrencies AMVP & 0.003 & 15 & 1461\\
        Dynamic - DJIA AMVP & -0.881 & 2 & 752\\
        Dynamic - Cryptocurrencies AMVP & 0.365 & 3 & 1461\\
        \hline
        3-Month T-Bill & 0.042 & $-$ & 752\\
        10-Year T-Bill & 0.046 & $-$ & 752\\
        \hline
        \hline
    \end{tabular}
    \caption{Static Measure of the Adaptive Minimum-Risk Rate from AMVP Efficient Frontier Progression}
    \label{tab:Static AMRR}
\end{table}

Theoretically, the AMVP aligns with standard principles of constrained optimization. The iterative introduction of synthetic assets and subsequent recalculation of the covariance structure systematically captures both first-order (mean returns) and second-order (covariance) effects, ensuring that convergence is both mathematically grounded and empirically observable.
The Adaptive Minimum-Risk Rate is the expected return of the emerging minimum-variance portfolio, illustrating the approach’s ability to harness optimal diversification and robust risk management throughout the convergence process. Table \ref{tab:Static AMRR} presents the static Adaptive Minimum-Risk Rate measure based on each portfolio examined in this paper. 

The value of the shadow rate for the AMVP when applied to the DJIA, which is 9.3\%, can be interpreted as the annualized return of the minimum-variance portfolio that is obtained after iteratively adding ten synthetic assets to the original DJIA asset pool. This relatively high shadow rate highlights that, under the historically observed returns for the constituents of the Dow Jones, the final AMVP retains sufficient exposure to higher-earning stocks while mitigating downside risk through diversification. The need for ten synthetic assets indicates that the portfolio converged after a moderate level of iterative refinement, with each synthetic asset capturing additional covariance information from previous iterations.

The annualized shadow rate of 0.3\% for the cryptocurrency portfolio reflects a much lower implied minimum-risk return in comparison to the DJIA. Despite strong episodes of appreciation, cryptocurrencies also exhibit heightened volatility, which can keep the minimal achievable risk–return trade-off relatively subdued. That the model required fifteen synthetic assets before stabilizing is a signal that larger fluctuations and complex interdependencies among cryptocurrencies necessitated more iterative steps to converge to a minimum-variance solution. 

Under the forward-looking regime—where ARFIMA-FIGARCH normal innovations are simulated via a NIG distribution—the shadow rate for the Dow Jones turns negative. Such an outcome can arise if the simulated return paths emphasize downside movements or periods of slow (or negative) growth, especially once heavy tails and long-memory features are incorporated.
A negative annualized rate does not necessarily imply that all returns are expected to be negative; rather, it indicates that within the model’s parametrization, the minimum-variance portfolio’s expected return is pushed below zero due to the combination of adverse simulated shocks, covariance structures, and potential negative drift. Only two synthetic assets were required to capture these effects adequately, reflecting relatively quick convergence under simulated data.

In contrast to the Dow Jones forward-looking scenario, the dynamic cryptocurrency AMVP exhibits a significantly higher annualized shadow rate (36.5\%). This suggests that once heavy tails and evolving correlations are included via NIG simulations, the algorithm still identifies pockets of high-return, albeit volatile, opportunities that push the minimum-risk return above historical baselines.
That the model converged after three synthetic assets indicates that although cryptocurrencies have high volatility, the inclusion of scenario-based extremes rapidly captures the potential for diversification under the fitted long-memory dynamics.

The 3-Month T-Bill (4.2\%) and 10-Year T-Bill (4.6\%) serve as proxies for near–risk-free returns, providing benchmarks against which the values of the Adaptive Minimum-Risk Rate can be compared. All covariance matrices in each iteration of the AMVP remain positive definite, thereby aligning with the theoretical foundations of the optimization method introduced in this paper. The range of values of the shadow rate highlights how historical and forward-looking methods can lead to widely divergent outcomes. Furthermore, the number of synthetic assets used before reaching convergence indicates the complexity of each asset universe, as more iterations are required when volatility and correlation structures are more intricate. Negative or unexpectedly low/high shadow rates in the dynamic setting reflect the model’s explicit incorporation of heavy-tailed distributions and evolving market conditions, yielding notably different portfolios than those identified under purely static historical estimates.

\subsection{The Adaptive Minimum-Risk Rate} \label{AMRR - main}

The Adaptive Minimum-Risk Rate (AMRR) is derived using the AMVP method, which iteratively constructs portfolios by introducing synthetic assets to progressively reduce portfolio variance. This dynamic process not only minimizes risk but also stabilizes the AMRR, providing a robust and data-driven alternative to standard measures of riskless rates. The construction of the AMRR is tightly linked to the iterative evolution of the efficient frontier, illustrating its convergence to the minimum-risk return. The optimization in appendix \ref{AMRR} ensures that the variance of the portfolio is minimized for each target return, under the constraints of weights being non-negative and the budget constraint being fully invested.

The AMVP method enhances the classical minimum-variance framework by introducing a dynamic, iterative mechanism that generates and incorporates synthetic assets. This approach captures the evolving interdependencies among the original assets and their combinations, allowing for a more adaptive and refined portfolio structure. A key feature of this iterative process is a feedback loop, where each newly created synthetic asset reflects the diversification properties identified in the preceding iteration, progressively improving portfolio efficiency. At each iteration, the expected return of the updated minimum-variance portfolio, referred to as the AMRR, is calculated as:

\begin{equation}
    \text{AMRR}_k = \mathbf{w}_k^\top \boldsymbol{\mu}_k
\end{equation}

\noindent where \(\boldsymbol{\mu}_k\) is the vector of mean-returns estimated from the updated dataset \(\mathbf{R}_k\). This quantity acts as a dynamic or ``shadow'' riskless rate in the absence of a conventional risk-free asset. Convergence is signaled by the stabilization of the variance of the MVP, $|\sigma_k^2 - \sigma_{k+1}^2| < \epsilon$, where \(\epsilon\) is a small tolerance level, indicating that further reductions in the variance are to be considered negligible. This dynamic mechanism ensures that the AMVP method adapts to changing market conditions while optimizing portfolio risk and return relationships.

Because the AMVP method relies on iterative updates of the covariance matrix, it readily adapts to both historical and forward-looking data regimes. For example, one may replace or augment \(\mathbf{R}_k\) with scenario-based returns generated using the normal innovations of the ARFIMA-FIGARCH model and subsequently transformed via a NIG distribution. Such an approach takes into account heavy-tailed distributions, skewness, and memory effects that might be overlooked by a purely historical analysis. At each iteration, the synthetic asset thus incorporate these dynamic distributional features, reflecting how extreme market events and shifting correlations reshape the risk profile.

In parallel with the iterative construction of synthetic assets, an efficient frontier is computed at each stage by solving a standard quadratic optimization problem over a range of target returns. Plotting these frontiers across successive iterations visually demonstrates an inward shift, signifying improved diversification and reduced variance for comparable return levels. The sequence of points \((\sigma_{\text{AMRR},k}, \text{AMRR}_k)\) illustrates the path of minimum-risk returns, culminating in a final stabilized portfolio once further variance improvements become negligible.

By progressively folding-in updated return and covariance information through newly created synthetic assets, the AMVP method  stands in contrast to static frameworks. Rather than relying on one-time estimates of means and covariances, it accounts for potential nonstationarities, structural breaks, and shifts in asset dependence. Each iteration’s synthetic asset effectively ``locks in'' the previously discovered risk-hedging benefits, leading to a portfolio that incorporates both historical and (optionally) model-based future dynamics.
As a result, the ultimate AMVP, along with its accompanying AMRR, serves as a robust indicator of the truly minimal variance achievable, given the available information and evolving market conditions.

\begin{figure}[h]
    \centering
    \includegraphics[width=1\linewidth]{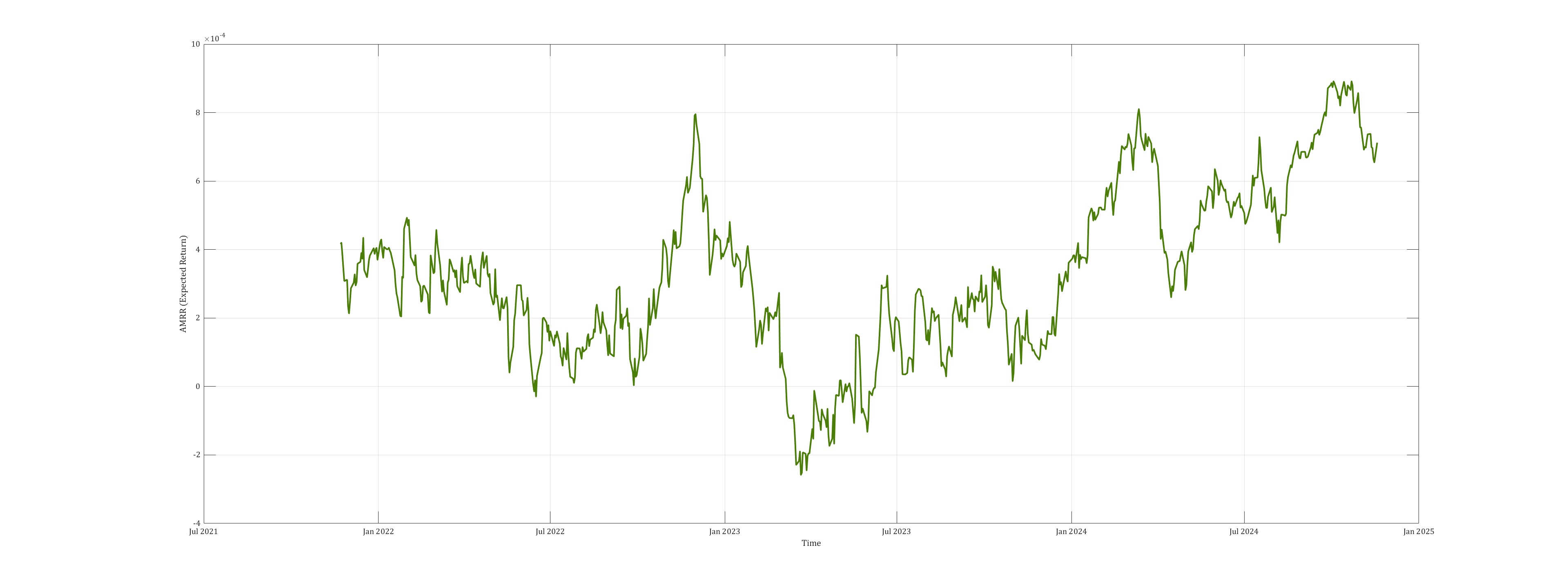}
    \caption{Adaptive Minimum-Risk Rate based on AMVP of Portfolio of Assets in DJIA}
    \label{fig:amrr dow}
\end{figure}

Figure \ref{fig:amrr dow} presents the time series of the AMRR for a portfolio of Dow Jones stocks, spanning the period from mid‐2021 through November 2024. Derived under the AMVP method, the AMRR at each point in time reflects the expected return of the minimum‐variance portfolio and serves as a ``shadow” risk‐free rate in the absence of a conventional risk‐free asset. By construction, the AMRR is computed from the mean returns vector \(\boldsymbol{\mu}_t\) and the portfolio weights \(\mathbf{w}_t\) associated with the minimum‐variance portfolio. This ensures that the AMRR is responsive to dynamic market conditions, including shifts in volatility, changes in correlation structures, and broader trends in equity prices. Accordingly, the AMRR provides a robust metric for capturing the evolution of the DJIA over the sample period.

To maintain temporal stability and ensure robustness in the estimation process, the AMRR is constructed using a rolling window with a window size of 252 trading days, corresponding to approximately one calendar year. This rolling estimation dynamically recalibrates the AMRR as new data points are incorporated while older observations are dropped. Such an approach ensures that the estimated parameters remain up-to-date with evolving market conditions, thereby minimizing the influence of outdated dynamics on the results. By adhering to this method, the time series of the AMRR is a more accurate representation of the underlying risk--return environment. This rolling-window framework is consistently applied across all AMRR time series constructed in this study.

There are periods in which the AMRR exhibits upward trends even though it pursues the lowest possible variance. Such trends typically correspond to phases of generally bullish market sentiment or periods of stable and improving corporate earnings. Conversely, declines in the AMRR reflect market conditions characterized by increased volatility on the part of the constituents of the  DJIA, deteriorating overall returns, or heightened asset correlations that constrain the diversification potential of the portfolio. Notably, instances where the AMRR dips below zero occur during extreme market stress, when negative shocks, low-yielding assets, or pronounced volatilities dominate the minimum‐variance portfolio, thereby pushing the implied ``safe” return to negative levels.

The dynamic behavior of the AMRR provides critical insights into the changing risk-return landscape of the DJIA. During periods of heightened market volatility or synchronized movements among its constituent stocks, the AMRR tends to decline, signaling that even the least risky portfolio cannot shield itself from adverse market conditions. In contrast, when market conditions improve—such as when asset returns exhibit greater dispersion that facilitates more effective diversification—the AMRR tends to rebound, signaling a more favorable environment for the construction of a risk-averse portfolio. Overall, this analysis highlights how fluctuations in the covariance structure and mean returns of the constituents of the DJIA directly influence the ``safe‐harbor” return implied by the adaptive minimum‐variance method, underscoring its dynamic adaptability and relevance to understanding evolving market conditions. 

\begin{figure}[h]
    \centering
    \includegraphics[width=1\linewidth]{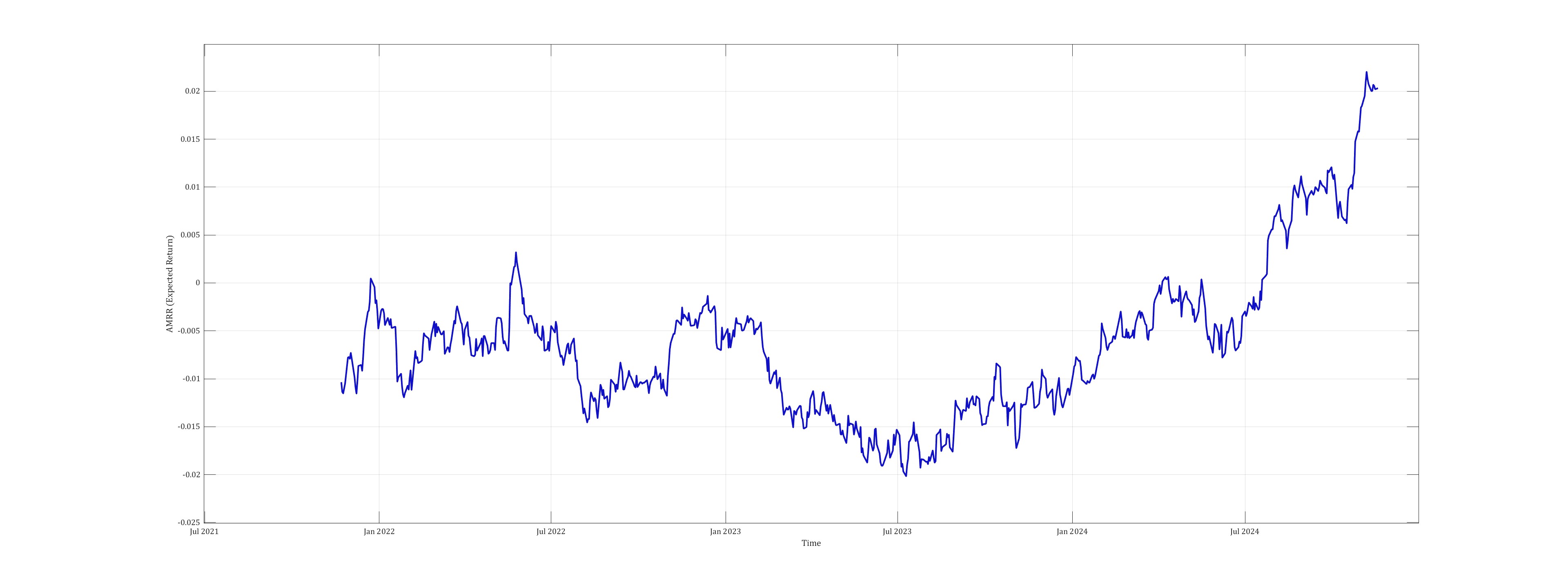}
    \caption{Adaptive Minimum-Risk Rate based on AMVP of \emph{forward-looking} Portfolio of Assets in DJIA}
    \label{fig:amrr ddow}
\end{figure}

In this regard, Figure \ref{fig:amrr ddow} shows the time series of the AMRR for the Dow Jones portfolio under forward-looking (dynamic) conditions, spanning the same time period as the other rates. Unlike the historical approach, these AMRR values arise from the data of a scenario generated via ARFIMA-FIGARCH normal innovations, subsequently transformed by a NIG distribution to accommodate heavy tails and evolving market dynamics. As before, each point on the plot represents the expected return of the minimum‐variance portfolio, interpreted here as a ``shadow'' risk‐free rate in the absence of a classical risk‐free instrument.

Notably, this dynamic method  can produce negative values of the AMRR during periods when the simulated downside risk dominates, reflecting the model’s emphasis on adverse shocks, high volatility, and unfavorable drift parameters. Over time, as the correlations, volatilities, and mean returns evolve in the simulated data, the AMRR adjusts accordingly, capturing new insights into the diversified risk landscape. The pronounced rise near the end of the sample period suggests that the forward-looking scenarios foresee an environment more favorable to risky assets, enabling the minimum‐variance allocation to yield positive expected returns despite its conservative orientation. Overall, this dynamic AMRR provides a window into how the Dow Jones universe might behave under heavy-tailed and long-memory conditions, extending beyond a purely backward‐looking analysis.

Over the plotted horizon, the time-varying AMRR reveals how forward-looking simulations can produce a markedly different risk–return profile than purely historical approaches. Downward trends often coincide with modelled crashes or heightened correlation episodes, whereas upturns emerge in simulations capturing surging asset prices or newly diversifying correlations. Consequently, this dynamic perspective offers a richer understanding of how portfolio risk and returns might evolve in volatile cryptocurrency markets, giving more weight to extreme outcomes and longer-run dependencies that traditional static methods may overlook.

The AMRR can be estimated using either backward-looking or forward-looking data, resulting in measures that often diverge in magnitude and temporal behavior. In the \emph{backward-looking} case, the AMRR is derived solely from historical asset returns, employing observed means and covariances. This approach reflects how the minimum-variance portfolio would have performed under actual past market conditions, making it highly intuitive for retrospective performance analyses and risk calibrations. By contrast, the \emph{forward-looking} AMRR arises from simulated or forecasted scenarios---such as those generated by ARFIMA-FIGARCH models with NIG innovations---that capture potential heavy tails, evolving correlation structures, and non-stationary volatility. Since these simulations embed hypothetical future market states, the forward-looking AMRR offers a risk--return perspective more attuned to unknown or extreme events, highlighting potential portfolio outcomes that may differ substantially from those implied by purely historical data.

Using both backward- and forward-looking measures is valuable in distinct economic and financial modeling contexts. Historical estimates inform stress testing, regulatory risk measures, and ex post performance audits, where the actual realized environment is of primary interest. Forward-looking estimates, on the other hand, support scenario planning, policy development, and strategic portfolio rebalancing in anticipation of possible market regimes. Integrating both measures allows practitioners and researchers to construct a more comprehensive picture of asset behavior, incorporating not only what has happened but also what could plausibly occur under alternative or adverse future states.

Beyond their interpretive power, these AMRR estimates possess several methodological advantages over other shadow riskless rate measures found in the literature. First, the AMRR is \emph{dynamically recalculated} whenever new synthetic assets are introduced, rather than relying on a fixed covariance structure. This iterative updating captures incremental diversification benefits discovered at each step, making the measure more robust to new information and evolving market conditions. Second, unlike high-dimensional PCA-based methods that can suffer from significant eigenvalue sensitivity, the AMRR method avoids such instabilities by operating directly on the portfolio’s covariance matrix and employing well-understood mean-variance principles. Lastly, the AMRR incorporates \emph{non-linear relationships} between the assets and real-world constraints, such as no short-selling, without resorting to layered approximations or factor models that may neglect key aspects of the underlying data. Consequently, the AMRR method provides a flexible, transparent, and empirically grounded way to estimate a time-varying risk-free proxy across a wide range of asset classes and market settings.

The AMRR provides a dynamic benchmark rate, improving upon static measures of the riskless rate. Its adaptability captures real-time market dynamics, offering valuable insights for macroeconomic modeling, asset pricing, and portfolio optimization. By dynamically adjusting to market conditions, the AMRR enhances the robustness of a financial model, making it a versatile tool for identifying systemic risks and optimizing asset allocations.

\subsection{Adaptive Minimum-CVaR Portfolio} \label{AMCVaRP}

AMCVaRP extends the AMVP method by replacing the variance criterion with a Conditional Value at Risk (CVaR) objective. We follow the method outlined in \citet{RockafellarUryasev2000} and \citet{McNeilFreyEmbrechts2015} to solve the CVaR optimization problem outlined in appendix \ref{AMCVARP}. Analogous to the AMRR in the AMVP context, the AMCVaRP method determines an Adaptive Min-CVaR Rate (AMCVarRR) at each iteration \(k\) to track expected returns under the current minimum-CVaR portfolio. In more detail, if \(\boldsymbol{\mu}_k\) is the mean return vector (computed over the same \(S\) scenarios in \(\mathbf{R}_k\)), then

\begin{equation}
    \text{AMCVarRR}_k
\;=\;
\mathbf{w}_k^{\star\,\top}
\boldsymbol{\mu}_k.
\end{equation}

This quantity serves as a dynamic ``shadow” rate that evolves alongside the portfolio as new synthetic assets are added and the risk measure (CVaR\(_{99}\)) is recalculated. By updating the optimization problem at each iteration, the AMCVaRP accommodates shifts in return distributions, covariance patterns, and extreme tail events.

\begin{figure}[h]
    \centering
    \begin{minipage}[t]{0.49\textwidth} 
        \centering
        \includegraphics[width=\textwidth]{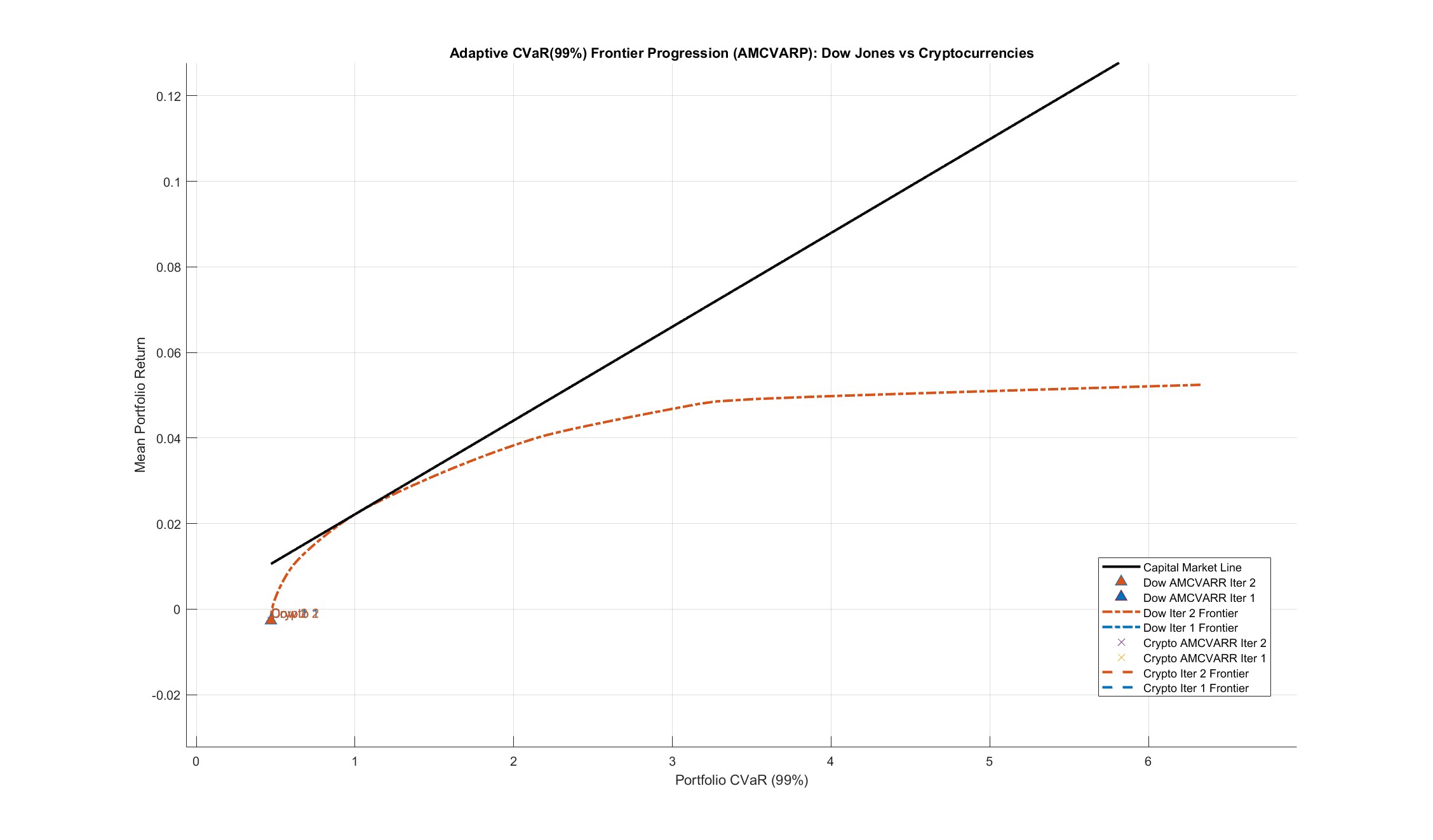} 
        \label{fig:image1}
    \end{minipage}
    \hfill
    \begin{minipage}[t]{0.49\textwidth} 
        \centering
        \includegraphics[width=\textwidth]{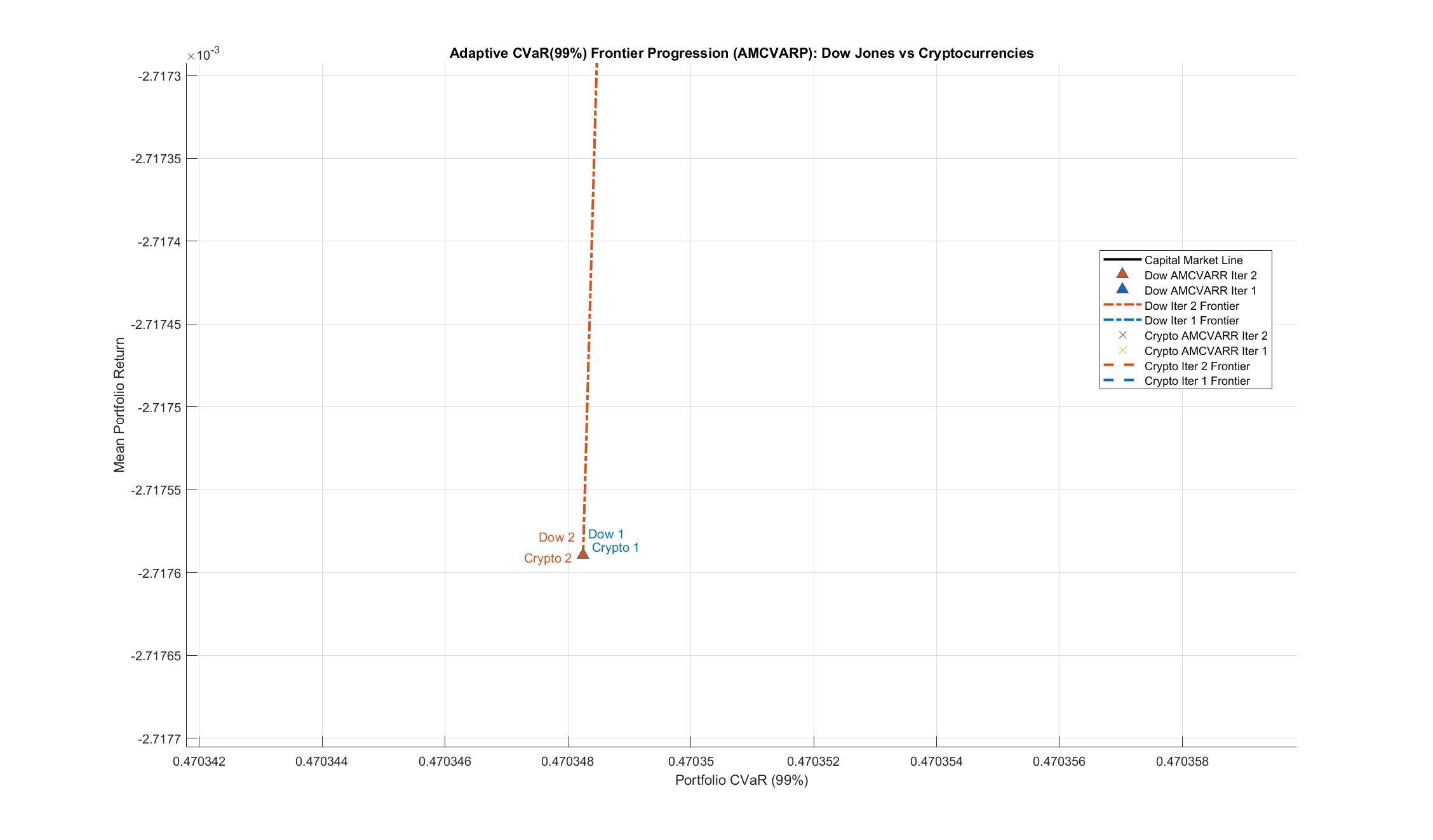} 
        \label{fig:image2}
    \end{minipage}
    \caption{AMCVaRP}
    \label{fig:AMCVARP}
\end{figure}

AMCVaRP extends the AMVP method by replacing the variance (or volatility) with the Conditional Value-at-Risk (CVaR) as the primary measure of risk. Under this formulation, the optimization problem aims to minimize the portfolio’s CVaR at a specified confidence level—here, 99\%. This approach follows the same iterative logic as the AMVP method, in which synthetic assets, derived from the weighted portfolio returns, are added to the asset universe at each iteration.

In this dynamic and forward-looking context, we generate 10,000 scenarios for both the DJIA and the cryptocurrencies by fitting an ARFIMA-FIGARCH model to historical returns and then transforming the normal innovations via a NIG distribution. Because NIG distributions can capture heavy tails and skewness, simulating 10,000 future paths allows a realistic approximation of extreme market outcomes.
By targeting $CVaR_{99}$, the AMCVaRP fully accounts for the far-left tail of the return distribution, effectively handling scenarios where portfolio losses might be exceptionally large. Indeed, $CVaR_{95}$ could miss part of this extreme behavior, especially in highly volatile or rapidly shifting markets.

In Figure \ref{fig:AMCVARP}, the left panel shows two sets of efficient frontiers, one for the DJIA and the other for the cryptocurrency portfolio, under the AMCVaRP method, each converging after two iterations. That is, each portfolio only requires two synthetic assets to reach its minimum CVaR solution under the 99\% criterion, illustrating that the incremental benefits of diversification from additional synthetic assets become negligible beyond the second iteration. The right panel zooms in to reveal the precise location of each portfolio’s ``Adaptive Min-CVaR Rate”—analogous to the AMRR from the AMVP method—across these two iterations. As with AMVP, this dynamic approach continually updates both the estimates of the return and the risk measure (CVaR), thereby reflecting evolving market conditions rather than relying on a static covariance matrix or single-shot sampling.

Modeling volatile market movements via the AMCVaRP is especially crucial when dealing with fat-tailed distributions, such as those that emerge from NIG scenarios in equity and cryptocurrency markets. Traditional mean-variance methods or narrower Value-at-Risk thresholds may underestimate the extent of extreme losses. By contrast, using $CVaR_{99}$ directly focuses on the worst 1\% of outcomes, enabling robust and proactive risk management in scenarios characterized by large price swings or other forms of tail-risk events. Furthermore, the alternate Capital Market Line (CML) in these figures is derived from the same forward-looking AMRR method discussed previously, now applied in the AMCVaRP setting. This yields a “baseline” or “reference” line emanating from the dynamic, risk-minimizing portfolio, rather than from a traditional (and often static) risk-free rate assumption.

Overall, the AMCVaRP method leverages an iterative, synthetic-asset expansion strategy and a CVaR-based objective to capture tail risk under a realistic generation of  heavy tailed distributions. That it converges within  two iterations underscores that, in many cases, only a modest number of synthetic assets may be required to lock in meaningful diversification benefits when a sufficiently large sample of forward-looking (NIG-based) returns is available. Such a method is therefore well-suited for complex or nonstationary markets, where ignoring tail risk can lead to underestimating extreme events and suboptimal risk mitigation.

\subsection{Analysis of Structural Breaks in the AMRR} \label{SB}

To find whether the time series of the AMRR for all portfolios including backward and forward-looking regimes capture the dynamics of the financial market, we conduct the Chow test to check for potential structural breaks. Econometrically, the Chow test identifies structural breaks by splitting the sample at potential breakpoints and comparing the fits of the regression models before and after the split. A statistically significant $p$-value suggests that the coefficients governing the relation between the dependent variable (here, the AMRR) and its explanatory factors has changed. This indicates a regime shift in the underlying data-generating process. Any possible structural breaks aligning with the events leading to market uncertainty implied by a volatility spike in the assets' expected returns would show that our measure of the shadow riskless rate can comprehensively capture the market risk and adjusts itself based on the dynamics of the market.
Section \ref{AMRR - main} introduced two measures of the AMRR or the shadow rate, and we now perform the Chow test on each time series to test for structural breaks. 

\begin{figure}[h]
    \centering
    \includegraphics[width=0.85\linewidth]{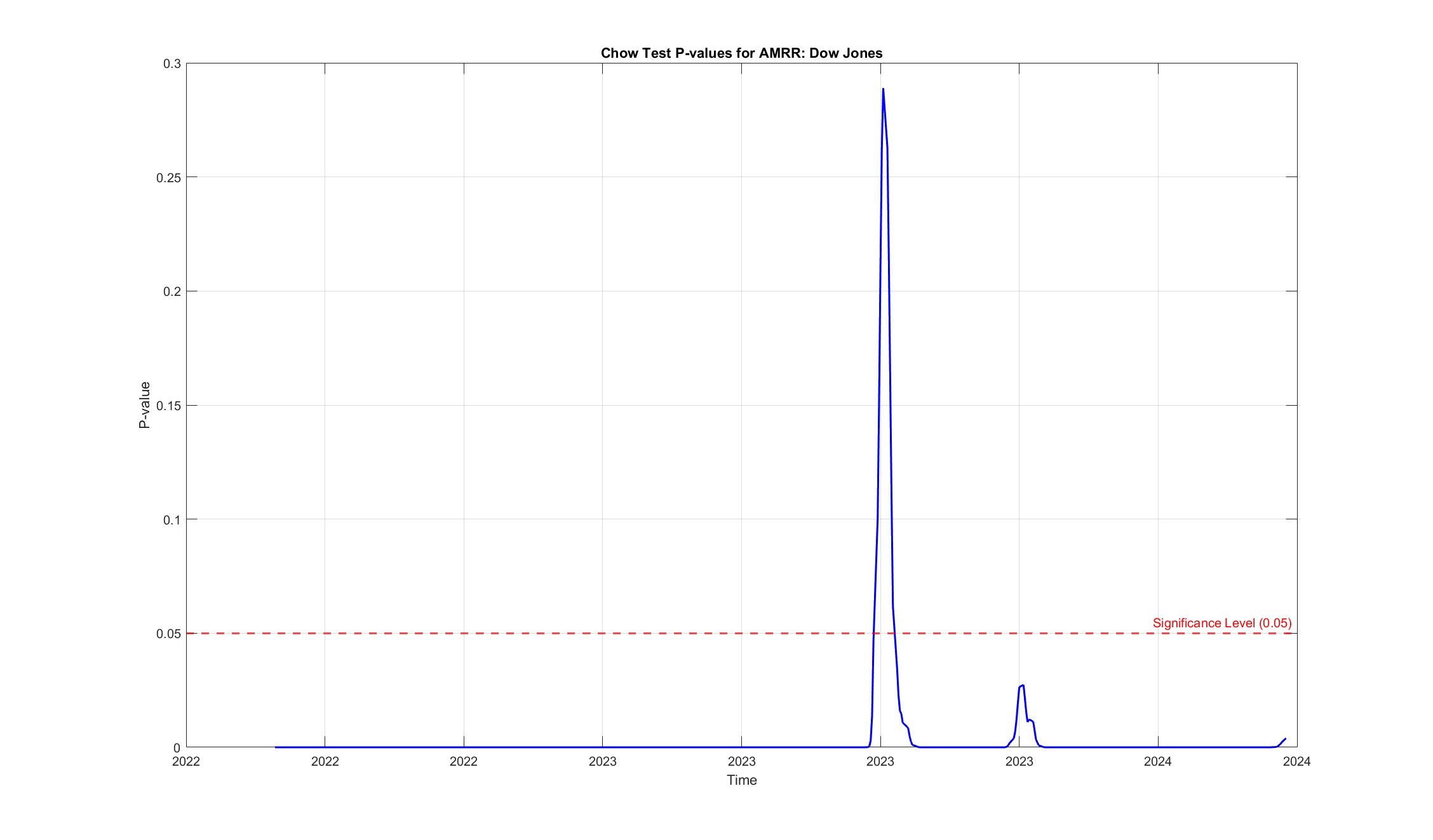}
    \caption{Chow Test for AMRR based on the Portfolio of Assets in the DJIA}
    \label{fig:chow dow}
\end{figure}

To begin with, Figure \ref{fig:chow dow} displays the $p$-values of the Chow test for structural breaks in the AMRR time series for the DJIA. The red dashed line marks the 5\% significance level ($p=0.05$), indicating the threshold below which the null hypothesis of no structural break is rejected. The significant structural break identified in early 2023 coincides with a critical shift in the behavior of the AMRR, signaling a potential change in the economic or financial environment affecting the constituents of the DJIA. 

In the context of the DJIA, the structural break observed in early 2023 can be linked to a notable period of market adjustment. This period saw increased uncertainty and volatility in financial markets due to global economic developments, including shifts in the monetary policy of the Federal Reserve and concerns about inflation. Specifically, early 2023 was characterized by the Federal Reserve’s aggressive rate hikes in response to persistent inflation, which created significant stress across equity markets. Higher interest rates typically depress equity valuations as the cost of borrowing rises and future cash flows are discounted at higher rates, leading to shifts in risk premia and investor expectations.

The AMRR, which dynamically reflects the expected returns of the minimum-risk portfolio, captures these shifts by responding to changes in the covariances of the assets and expected returns. During this period, the increased dispersion in the returns of the constituents of the DJIA, driven by their varied sensitivities to interest rates and macroeconomic shocks, probably led to abrupt changes in the AMRR's behavior. This abrupt adjustment is detected as a structural break by the Chow test, aligning with theoretical predictions that economic or policy shocks can alter the risk--return trade-off in equity markets.

Furthermore, the post-2023 period appears to feature residual adjustments in the AMRR, as the $p$-values temporarily drop near the significance threshold again later in 2023. This secondary signal may reflect a market stabilization following the Federal Reserve’s communication of a terminal rate or an emerging clarity about inflation trajectories. These observations are consistent with the financial literature, which highlights how structural breaks often coincide with major policy changes or with market stress, underscoring their relevance for dynamic portfolio optimization and risk management strategies.

The application of the Chow test in this context demonstrates the importance of detecting regime shifts in financial econometrics. Structural breaks in the AMRR provide a forward-looking signal of changing market conditions, allowing investors and policymakers to adjust strategies in response to evolving risk--return dynamics. This method’s ability to identify such breaks strengthens its utility in the analysis of financial time series, particularly for portfolios exposed to systemic risks.

\begin{figure}[h]
    \centering
    \includegraphics[width=1\linewidth]{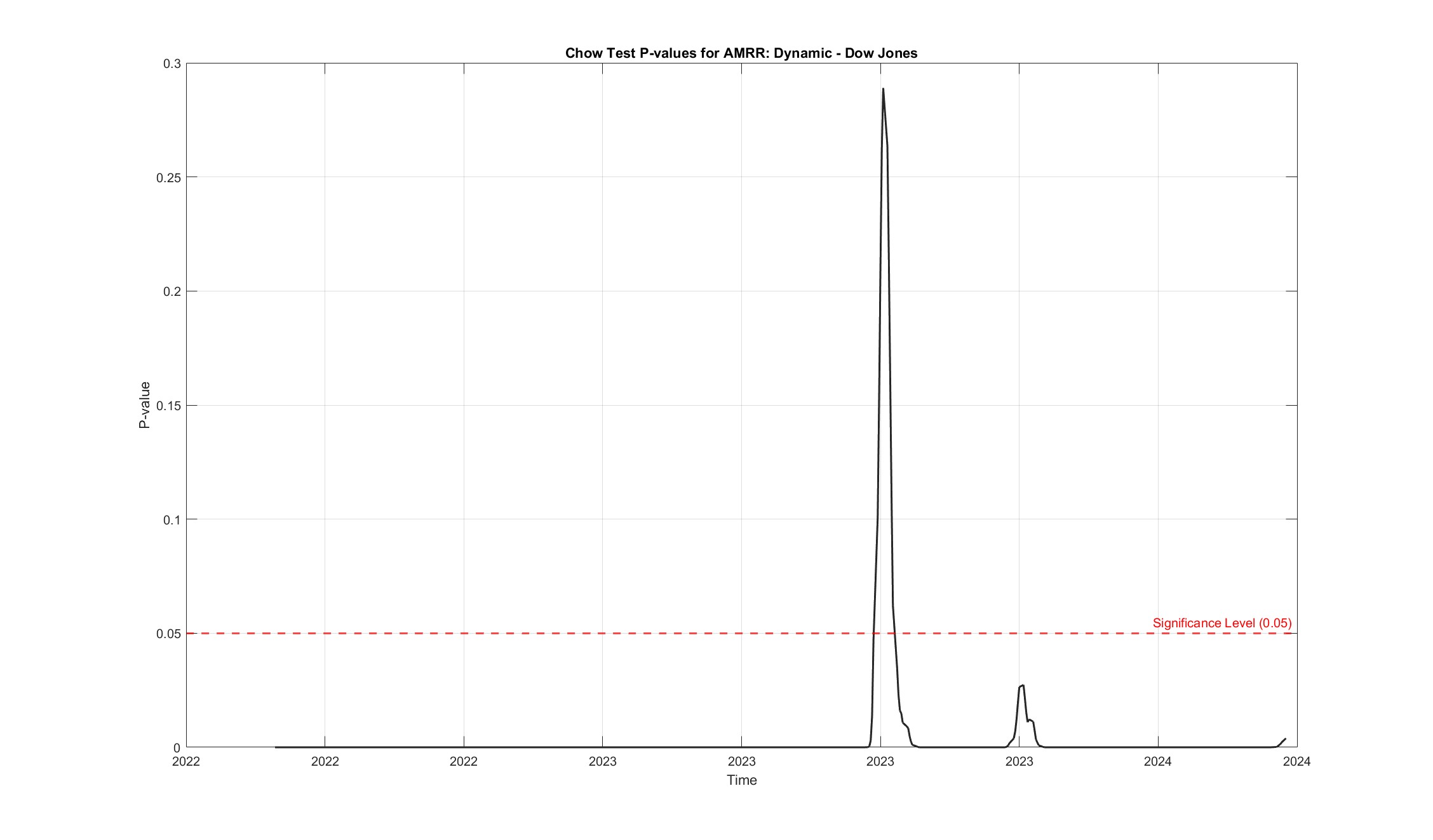}
    \caption{Chow Test for foward-looking AMRR based on portfolio of DJIA}
    \label{fig:chow ddow}
\end{figure}

Second, we examine the ex-post measure of the AMRR time series to determine whether the market dynamics modeled using a heavy-tailed distribution can capture significant structural breaks. Figure \ref{fig:chow ddow} displays the $p$-values of the Chow test AMRR derived from the forward-looking measure of the DJIA portfolio. A structural break is identified in early 2023, aligning closely with the results obtained from the Chow test applied to the backward-looking measure of the AMRR for the DJIA portfolio. The similarity in the timing of the structural breaks identified by the Chow test for both forward-looking and backward-looking measures highlights the robustness of the modeling approach. 
The forward-looking AMRR is based on NIG distribution scenarios that are designed to mimic the distribution of asset returns, which are heavy-tailed and negatively skewed, characteristics commonly observed in financial markets. The ability of the forward-looking model to replicate these statistical properties ensures that it captures extreme market dynamics, including the conditions that led to the structural break in early 2023.

That structural break corresponds to a period of significant market turbulence. As noted earlier, that period was marked by an aggressive tightening of monetary policy by the Federal Reserve in response to persistent inflationary pressures. The resulting increase in interest rates led to a broad revaluation of asset prices, particularly in sectors sensitive to interest rates, which are prominently represented in the DJIA. The heavy-tailed NIG scenarios effectively capture the large deviations in asset returns associated with such market stress, allowing the forward-looking AMRR to signal the structural break at a similar time as the backward-looking one.

While the backward-looking measure reflects realized data, the forward-looking measure incorporates the possibility of extreme market outcomes, providing a more comprehensive view of risk. The alignment of the structural breaks identified by both approaches validates the ability of the forward-looking method to model the true underlying dynamics of asset returns. Moreover, the use of NIG-based scenarios adds an additional layer of robustness by taking into accoun non-linear dependencies and extreme risk events. The similarity in the timing of structural breaks emphasizes that the forward-looking measure not only predicts potential future risks but also aligns closely with the ex-post observations of market behavior, enhancing its utility for risk management and dynamic portfolio optimization. 

Cryptocurrencies, unlike equities, are subject to idiosyncratic risks such as regulatory announcements, technological shifts, and speculative behavior, which create abrupt and significant changes in market conditions. The Chow test's ability to detect a structural break at a different time than the backward-looking measure underscores the importance of using forward-looking scenarios for the risk assessment in highly volatile markets. It highlights the value of forward-looking measures for identifying early signals of regime shifts in markets characterized by extreme volatility. While ex-ante measures provide insights based on realized data, ex-post measures extend this analysis by anticipating potential tail-risk events. For cryptocurrencies, which exhibit high sensitivity to unique market dynamics, the forward-looking AMRR offers a robust tool for dynamic portfolio optimization and risk management. The divergence in the timing of the structural breaks demonstrates the complementary nature of the two approaches, emphasizing the importance of employing both measures to obtain a comprehensive understanding of risk--return dynamics in classes of highly volatile assets.

\section{Results} \label{results}

The results of this study underscore the robustness of the Adaptive Minimum Variance Portfolio and the Adaptive Minimum Risk Rate methods in capturing nonlinear dependencies and higher-order moments in asset return data that traditional portfolio optimization methods often overlook. By iteratively introducing synthetic assets into the portfolio until the least achievable variance is obtained, the AMVP method comprehensively addresses the non-Gaussian characteristics of asset returns. Unlike standard mean-variance optimization, which assumes normally distributed returns and linear relations between the assets, the AMVP method acknowledges and incorporates the empirical realities of heavy tails, skewness, and nonlinear interdependencies, as commonly observed in real-world financial markets. This iterative process ensures that the portfolio weights reflect the evolving dynamics of risk and return while adapting to changing correlation structures and volatility regimes.

The Hurst exponent, estimated as \(H = 0.5 + d(v)\), provides additional evidence of LRD in the constructed portfolios. In the standard ARFIMA(0, \(d\), 0) framework, a fractional parameter \(d \ge 0.5\) often implies that the process is no longer stationary. However, in the context of modeling volatility (e.g., using FIGARCH or FIEGARCH), fractional parameters above 0.5 can remain weakly stationary in terms of conditional variance, depending on the specific structure of the recursion. In practical terms, a fractional parameter \(d(v)\) near or above 0.5 suggests a hyperbolic decay in the autocorrelation function of the volatility, leading to prolonged clustering in the volatility and a slower dissipation of shocks.

\begin{table}[h]
    \centering
    \begin{tabular}{ccc}
    \hline
    \hline
       Shadow Rate (AMRR)  & $d(v)$ & Hurst Exponent (\textit{H})\\
    \hline
       DJIA  & 0.8668 & 1.3668 \\
       Cryptocurrencies  & 0.3651 & 0.8651 \\
       Dynamic DJIA & 0.4380 & 0.9380 \\
       Dynamic Cryptocurrencies  & 0.4043 & 0.9043 \\
       \hline
       10-Year Treasury Bill & 0.9999 & 1.4999 \\
       3-Month Treasury Bill & 0.9939 & 1.4939 \\
    \hline
    \hline
    \end{tabular}
    \caption{LRD in the log-transformed AMRR Time Series}
    \label{tab:LRD}
\end{table}

Table \(\ref{tab:LRD}\) presents the persistence in volatility parameters, \(d(v)\), and the corresponding Hurst exponents, \(H\), for the log-transformed AMRR time series across portfolios of equities (DJIA), cryptocurrencies, and Treasury bills. These estimates quantify the degree of LRD in the volatility dynamics and shed light on the stability and risk--return profiles of the AMRR, which is derived under the AMVP method. The parameter \(d(v)\) captures the fractional integration in volatility and thus reflects how shocks persist over time. A value of \(d(v) > 0.5\) indicates the presence of long-memory effects, whereby volatility shocks decay hyperbolically rather than exponentially, as is typical of short-memory processes.

For the AMRR derived from the DJIA portfolio, the value of \(d(v) = 0.8668\) highlights the considerable long-memory effects in equity market volatility, aligning with a vast body of empirical work indicating that stock market volatility clusters over extended periods due to macroeconomic cycles, corporate earnings trends, and propagation of systematic risk. In contrast, the cryptocurrency portfolio has \(d(v) = 0.3651\), indicating weaker long-memory effects in its volatility dynamics. This lower persistence resonates with the unique nature of cryptocurrency markets, where volatility is driven more by episodic shocks---such as regulatory announcements or speculative trading---than by persistent economic factors. Dynamic AMRR measures constructed via forward-looking NIG scenarios reduce \(d(v)\) relative to these backward-looking estimates (\(d(v) = 0.4380\) for DJIA and \(d(v) = 0.4043\) for cryptocurrencies), indicating that incorporating scenario-based models and heavy-tail considerations dampens any overreliance on entrenched historical patterns.

The persistence in Treasury bills, indicated by \(d(v) = 0.9999\) for both the 10-year and 3-month maturities, suggests nearly unit-root dynamics in their volatility processes. This extreme persistence stems from the fact that Treasury bills, as traditional hedging instruments, are more insulated from market-specific shocks, thereby enjoying heightened predictability and stability compared to risky assets. Consistent with these \(d(v)\) estimates, the Hurst exponent \(H\) likewise indicates strong persistence in volatility dynamics when \(H > 0.5\). For the DJIA AMRR, \(H = 1.3668\) confirms pronounced volatility persistence, whereas \(H = 0.8651\) for cryptocurrencies aligns with weaker, yet still evident, long-memory effects. The dynamic AMRR estimates for both DJIA and cryptocurrencies have lower Hurst exponents (\(H = 0.9380\) and \(H = 0.9043\), respectively), in line with the reduced reliance on historically ingrained patterns once forward-looking scenarios are incorporated.

Treasury bills have the highest Hurst exponents, with \(H = 1.4939\) for 3-month bills and \(H = 1.4999\) for 10-year notes, reflecting the extreme persistence and smoothness that one typically associates with risk-free assets. These findings illustrate why Treasury bills are often viewed as the cornerstone of conventional hedging strategies. By contrast, the AMRR provides a ``shadow rate'' constructed from portfolios entirely made up of risky assets. Although the AMRR exhibits lower persistence in volatility compared to Treasury bills, it achieves a comparable or sometimes higher expected return precisely because it is formed through the AMVP optimization process. This novel contribution---achieving a shadow rate with a higher expected return at lower long-range dependence than many traditional riskless benchmarks---underscores the significance of the minimum-risk rate in environments where conventional risk-free instruments may be inaccessible, illiquid, or otherwise impractical.

In extending this perspective, an alternative Capital Market Line (CML) can be constructed using the shadow risk-free rate as the intercept. This CML captures the trade-off between risk and return in settings where a conventional risk-free asset is either absent or unsuitable. By reflecting the nonlinear dependencies, non-Gaussian distributions of the returns, and time-varying market conditions underlying the AMVP method, this new CML provides a more nuanced and realistic depiction of investment opportunity sets. In particular, investors operating in high-volatility or markets for unconventional assets---such as cryptocurrencies---may benefit from a baseline that appropriately takes into account heavy tails, skewness, and evolving correlation structures. In this regard, the AMRR redefines the notion of a ``risk-free rate'' by situating it within the portfolio itself, thereby allowing for endogenous hedging mechanisms that can adapt as the portfolio's composition or market conditions evolve.

It is qualified by the word ``shadow'' because it emerges from a portfolio of \(n + 1\) assets, where the extra asset is a synthetic linear combination of the initial \(n\), so this rate cannot be separated from the portfolio’s overall structure. Unlike a conventional risk-free rate, which is presumably independent of other holdings, the shadow rate’s construction relies upon residual terms that amplify diversification benefits. These terms make it impossible to isolate the synthetic asset without forfeiting the reduction in the variance inherent in the AMVP method. From a scientific standpoint, this indivisibility is a natural consequence of solving for the lowest possible variance: the residual terms reflect combined sources of diversification that transcend the simpler additive structure of standard mean-variance models.

The residual terms embedded in the synthetic asset allow the portfolio to achieve lower minimum variance than would be possible under traditional optimization paradigms. Their inclusion enhances the portfolio’s capacity to exploit intricate market dependencies and to mitigate exposure to extreme tail events. This property is particularly beneficial in markets dominated by nonlinearities and regime shifts---such as in cryptocurrencies---where conventional mean-variance approaches often fail to capture the true extent of risk concentration. The AMVP method’s ability to reveal---and capitalize on---these hidden correlations and tail risks not only reduces the overall variance but also fosters more accurate risk-adjusted performance metrics.

The implications of this work resonate in both theoretical and applied finance. From a theoretical standpoint, the shadow rate broadens the scope of models like Black’s Zero-Beta CAPM by providing a genuine proxy for the risk-free rate in circumstances where no such asset is explicitly available. Bridging this gap between theory and empirical practice paves the way for more refined asset pricing models that acknowledge the reality of heavy tails, skewness, and evolving market correlations. Practically, investors gain access to novel instruments for hedging and constructing portfolios, through the alternative CML and shadow risk-free rate. Indeed, these constructs enable investors to treat volatility as an asset class---mirroring how the VIX is tradeable---due to the predictable patterns in LRD-based volatility clustering. By extending this to portfolios of the DJIA and cryptocurrencies, the AMRR provides a platform for creating portfolio-specific volatility indices. Such indices can serve as tradeable instruments, offering robust hedging opportunities against market turbulence and further illustrating the feasibility of volatility-based products in diverse market settings.

A key insight emerging from the AMRR method lies in its capacity to model and predict volatility by explicitly incorporating the long-range dependence inherent in financial time series. The persistent positive values of \(d(v)\) in the AMRR imply that there are slowly decaying shocks in volatility, reinforcing the notion that volatility---while stochastic---remains predictable over extended horizons. This predictability underpins the tradability of volatility indices such as the VIX and can similarly support a portfolio-specific volatility index based on AMRR. When further combined with the shadow risk-free instruments derived from the AMVP method, which themselves exhibit a higher expected return at lower long-range dependence relative to many conventional riskless assets, the practitioner gains a powerful suite of hedging tools. This integration of LRD-based predictability and endogenous risk-free rates constitutes a novel frontier in asset pricing and portfolio management, particularly for kinds of asset classes where traditional hedging mechanisms are unavailable. Consequently, the AMVP and AMRR methods advance both the theoretical underpinnings of modern portfolio theory and its practical application, offering more precise risk-adjusted returns and paving the way for broader innovations in risk management and asset allocation.

\section{Conclusion} \label{conclusion}

This paper introduces the Adaptive Minimum-Variance Portfolio (AMVP), Adaptive Minimum-Risk Rate (AMRR), and Adaptive Minimum-CVaR Portfolio (AMCVaRP) frameworks to address critical challenges in portfolio optimization and asset pricing. These methods advance traditional models by incorporating nonlinear dependencies, non-Gaussian return distributions, and long-range dependence into portfolio construction. Using empirical data from DJIA stocks and cryptocurrencies, the study demonstrates the efficacy of these approaches in capturing heavy tails, skewness, and evolving market conditions.

The AMVP method iteratively integrates synthetic assets into portfolios, achieving configurations with minimal variance while accounting for nonlinear dependencies. The AMRR acts as a dynamic shadow risk-free rate, adapting to changing market conditions and offering an alternative to traditional risk-free assets. Additionally, the AMCVaRP optimizes portfolios under extreme downside risks, enhancing resilience in turbulent markets. These methods collectively improve risk management and portfolio diversification, enabling the construction of alternative Capital Market Lines and tradeable risk metrics like volatility indices.

Empirical findings further emphasize the adaptability of these frameworks in identifying and responding to structural breaks in financial markets. By applying forward-looking Chow tests on AMRR time series, we identified significant market events and shifts in volatility regimes, showcasing the predictive capacity of these methods. These insights are particularly valuable for volatile asset classes like cryptocurrencies, which exhibit distinct market dynamics compared to equities. The ability to capture such shifts highlights the robustness of the proposed methodologies across diverse market environments.

From a theoretical perspective, the AMRR bridges the gap between classical portfolio theory and modern, data-driven approaches, with potential applications in asset pricing models like Black’s Zero-Beta CAPM. Practically, these methods provide investors and policymakers with innovative tools to model volatility, enhance diversification, and detect structural breaks aligned with major market events. Future research should expand these frameworks to multi-asset portfolios, explore their scalability across various asset classes, and assess their implications for financial stability and macroeconomic policy.

\newpage
\bibliography{output}

\pagebreak
\appendix
\section*{Appendix}

\section{ARFIMA(1, $d(m)$, 1)-FIGARCH(1, $d(v)$, 1)} \label{ARFIMA}

Long-memory effects in the mean, as captured by ARFIMA, indicate that historical returns exert an influence over future values for extended periods. Analogously, when volatility exhibits long memory, as modeled by FIGARCH, episodes of pronounced market fluctuations tend to persist. To incorporate both of these features, we specify the arithmetic returns \(z_{t}\) using the ARFIMA(1, \(d(m)\), 1) process:
\[
\phi(L)\,(1 - L)^{d(m)} z_{t} \;=\; \theta(L)\,\varepsilon_{t},
\]
where \(d(m)\) represents the fractional differencing parameter governing long memory in the mean, and \(\varepsilon_{t}\) is a white noise innovation. For volatility, we employ a FIGARCH(1, \(d(v)\), 1) framework:
\[
\phi(L)\,(1 - L)^{d(v)}\,\varepsilon_{t}^{2} \;=\; \omega \;+\; [1 - \beta(L)]\,\nu_{t},
\]
in which the parameter \(d(v)\) measures the degree of long-memory dependence in volatility and $\nu_{t}$ is the normal innovations term.

\section{Adaptive Minimum-Variance Portfolio} \label{AAMVP}

The MVP is obtained by solving
\begin{equation}
    \min_{\mathbf{w}_t} \, \mathbf{w}_t^\top \boldsymbol{\Sigma}_t \mathbf{w}_t
\quad \text{subject to} \quad \mathbf{w}_t^\top \mathbf{1} = 1, \quad \mathbf{w}_t \geq 0,
\end{equation}
where \(\mathbf{1}\) is an \(N \times 1\) vector of ones, ensuring the portfolio is fully invested and prohibiting short positions. Provided \(\boldsymbol{\Sigma}_t\) is invertible, the solution for the MVP weights is
\begin{equation}
    \mathbf{w}_t 
= \frac{\boldsymbol{\Sigma}_t^{-1}\mathbf{1}}{\mathbf{1}^\top \boldsymbol{\Sigma}_t^{-1}\mathbf{1}},
\end{equation}
which yields a portfolio variance
\begin{equation}
    \sigma_t^2 
= \mathbf{w}_t^\top \boldsymbol{\Sigma}_t \mathbf{w}_t.
\end{equation}

 at iteration \(k\), we compute
\[
\mathbf{w}_k 
= \frac{\boldsymbol{\Sigma}_k^{-1}\mathbf{1}}{\mathbf{1}^\top \boldsymbol{\Sigma}_k^{-1}\mathbf{1}},
\]
then form the synthetic return series
\[
r_{\text{AMVP},k} 
= \mathbf{R}_k\, \mathbf{w}_k,
\]
which is appended to the existing dataset:
\[
\mathbf{R}_{k+1} 
= \big[\mathbf{R}_k, \; r_{\text{AMVP},k}\big].
\]
We update the covariance matrix \(\boldsymbol{\Sigma}_{k+1}\) accordingly:
\[
\boldsymbol{\Sigma}_{k+1}
= \frac{1}{T-1}\, \big(\mathbf{R}_{k+1} - \bar{\mathbf{R}}_{k+1}\big)^\top 
                      \big(\mathbf{R}_{k+1} - \bar{\mathbf{R}}_{k+1}\big),
\]
where \(\bar{\mathbf{R}}_{k+1}\) is the mean return vector over \(\mathbf{R}_{k+1}\). We then solve for the next iteration's MVP, thus generating an increasingly augmented asset space at each step. The process continues until the variance fails to show significant improvement, which we enforce via a convergence criterion
\[
\bigl|\sigma_k^2 - \sigma_{k+1}^2\bigr| < \epsilon,
\]
for a suitably small tolerance \(\epsilon\). Through this iterative construction, the Adaptive Minimum Variance Portfolio (AMVP) systematically seeks new synthetic assets that contribute to variance reduction, and thereby yields a dynamic portfolio solution extending beyond the conventional mean-variance framework.

\section{Adaptive Minimum-Risk Rate} \label{AMRR}

Section \ref{AMRR - main} mentions the use of rolling window estimation to construct the time series of AMRR. We follow \citet{Hamilton1994} to estimate model parameters dynamically as new data becomes available. 

Let the time series data be \(\{y_t\}_{t=1}^T\), where \(T\) is the total number of observations. Define the rolling window size as \(P\), with \(P < T\). The \(k\)-th window consists of observations:
\[
\mathcal{P}_k = \{y_k, y_{k+1}, \ldots, y_{k+P-1}\}, \quad k = 1, 2, \ldots, T-P+1.
\]

For each window \(\mathcal{P}_k\), estimate the parameters of a model \(f(y_t, \boldsymbol{\theta})\), where \(\boldsymbol{\theta}_k\) represents the parameter vector estimated for the \(k\)-th window:

\begin{equation}
\hat{\boldsymbol{\theta}}_k = \arg\min_{\boldsymbol{\theta}} \sum_{t=k}^{k+P-1} \ell(y_t, f(y_t, \boldsymbol{\theta})),
\end{equation}

where \(\ell(\cdot)\) is the loss function, often the squared error for regression models or the log-likelihood for probabilistic models. The result is a time series of parameter estimates:
\[
\{\hat{\boldsymbol{\theta}}_1, \hat{\boldsymbol{\theta}}_2, \ldots, \hat{\boldsymbol{\theta}}_{T-P+1}\}.
\]
These estimates provide insights into how the model parameters evolve over time. Using the estimated parameters \(\hat{\boldsymbol{\theta}}_k\), rolling predictions for the dependent variable \(y_t\) can be made:

\begin{equation}
\hat{y}_t^{(k)} = f(y_t, \hat{\boldsymbol{\theta}}_k), \quad t \in \mathcal{P}_k.
\end{equation}

Eq. (7) is resulting time series over k-horizons as mentioned in eq. (1) in section \ref{AMRR - main}.

\begin{figure}[h]
    \centering
    \includegraphics[width=1\linewidth]{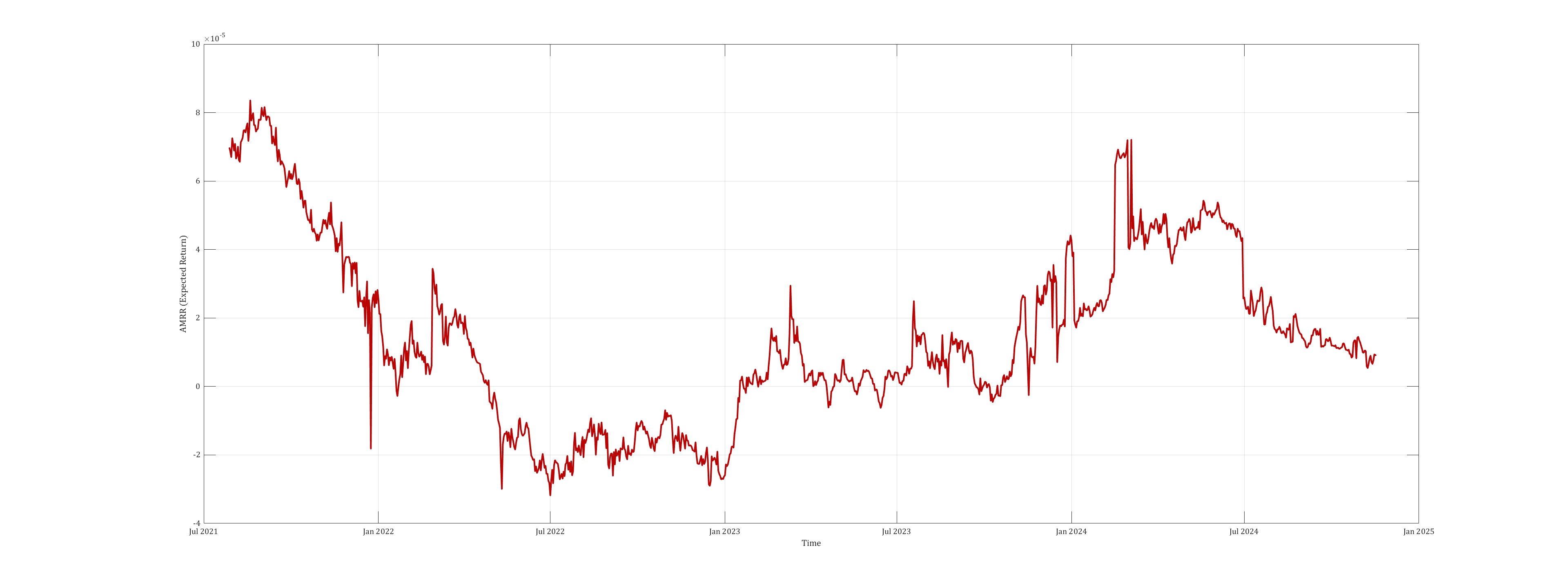}
    \caption{Adaptive Minimum-Risk Rate based on AMVP of Portfolio of Cryptocurrencies. A rise in the AMRR indicates that the minimum‐variance configuration is capturing a higher expected return despite aiming for the lowest possible variance. Such increases may coincide with market upswings or emergent demand for specific digital currencies. Conversely, a decline in the AMRR can reflect heightened volatility in the crypto market, diminished return prospects, or intensified correlations that constrain diversification. In instances where the AMRR drops below zero, it suggests that negative shocks, extremely low or negative returns, and high volatility dominate the risk–return landscape, making it challenging for a strictly minimal‐variance strategy to maintain positive performance.}
    \label{fig:amrr crypto}
\end{figure}

\begin{figure}[h]
    \centering
    \includegraphics[width=1\linewidth]{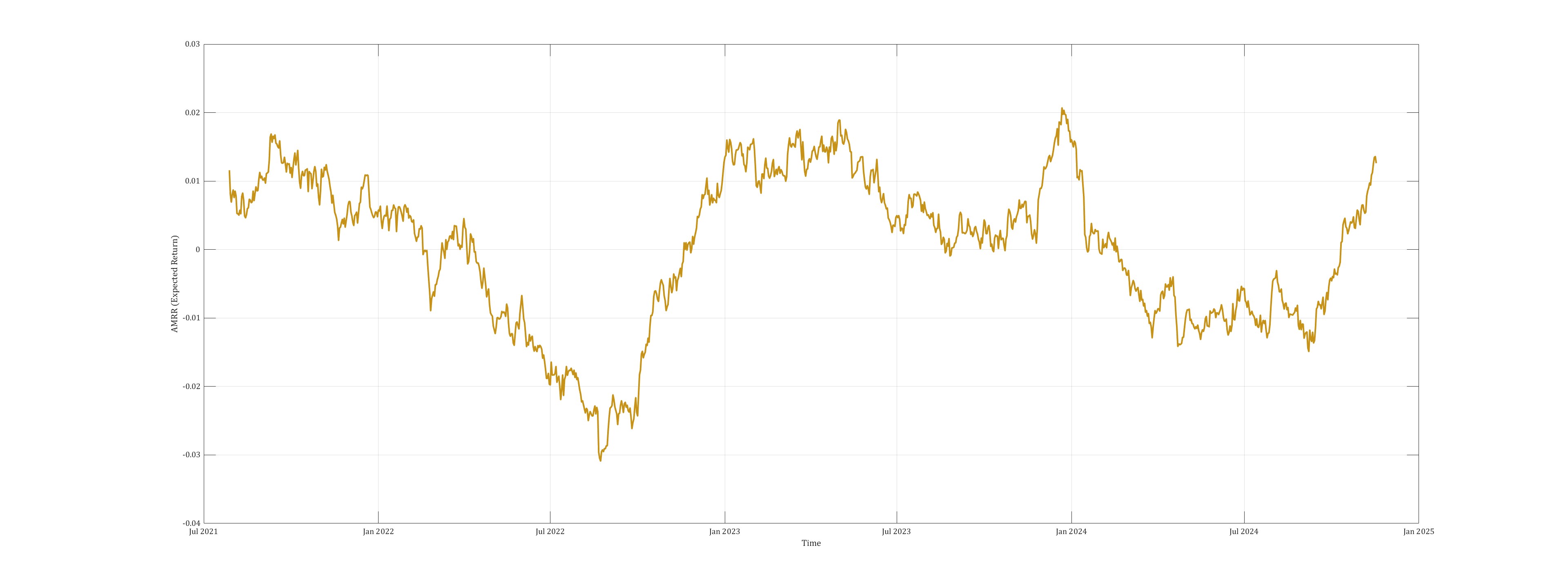}
    \caption{Adaptive Minimum-Risk Rate based on AMVP of \emph{forward-looking} Portfolio of Cryptocurrencies. Periods where the AMRR moves deeper into negative territory suggest episodes of simulated downside shocks or heightened volatility within the digital asset universe, driving the minimum-variance configuration toward negative return expectations. Conversely, the rebound and positive swings of the AMRR correspond to scenarios that highlight more favorable conditions for select cryptocurrencies, thereby boosting the portfolio’s expected return even when variance is minimized. Such variations underscore the sensitivity of the AMRR to both the structural properties of the simulation model and the underlying market fundamentals that shape digital asset performance.}
    \label{fig:amrr dcrypto}
\end{figure}

\pagebreak

\section{Adaptive Minimum-CVaR Portfolio} \label{AMCVARP}

Consider a set of \(S\) return scenarios for \(N_k\) assets at iteration \(k\), captured in the matrix \(\mathbf{R}_k \in \mathbb{R}^{S \times N_k}\). Each row \(\mathbf{R}_k[s,:]\) corresponds to the returns of all assets under scenario \(s\). A portfolio weight vector \(\mathbf{w}_k \in \mathbb{R}^{N_k}\) must satisfy the constraints 
\[
\mathbf{w}_k \geq \mathbf{0}, 
\quad 
\mathbf{w}_k^\top \mathbf{1} = 1,
\]
reflecting non-negativity and full investment. Under scenario \(s\), the portfolio’s realized return (or loss if negative) is given by 
\[
r_{s,k} \;=\; \mathbf{R}_k[s,:] \,\mathbf{w}_k.
\]

To incorporate CVaR at the \(\alpha\) level (e.g., \(\alpha = 0.99\)), the optimization problem can be formulated using an auxiliary variable \(t\) and slack variables \(\{z_s\}_{s=1}^S\). In one common linear-programming representation, minimizing CVaR translates into minimizing
\[
t 
\;+\;
\frac{1}{(1-\alpha)\,S} \sum_{s=1}^S z_s,
\]
subject to
\[
z_s \;\ge\; -\,r_{s,k} \;-\; t, 
\quad
z_s \;\ge\; 0 
\quad\text{for all } s=1, \ldots, S,
\]
where \(-\,r_{s,k}\) represents a loss if \(r_{s,k}\) is negative. The decision variables \(\mathbf{w}_k\), \(t\), and \(\{z_s\}\) are all solved simultaneously, yielding the optimal minimum-CVaR portfolio \(\mathbf{w}_k^{\star}\). Because \(\mathbf{w}_k\) must also satisfy non-negativity and the budget constraint, the final optimization system becomes
\[
\begin{aligned}
\min_{\substack{\mathbf{w}_k,\,t,\,\{z_s\}}} 
&\;
t 
\;+\;
\frac{1}{(1-\alpha)\,S} \sum_{s=1}^S z_s,\\
\text{subject to}\;\;
& z_s \;\ge\; -\,r_{s,k} \;-\; t, \;\; z_s \;\ge\; 0, \quad s=1,\ldots,S,\\
& \mathbf{w}_k \;\ge\; \mathbf{0}, 
\quad
\mathbf{w}_k^\top \mathbf{1} \;=\; 1.
\end{aligned}
\]
Upon solving this problem, the resulting portfolio \(\mathbf{w}_k^{\star}\) achieves the lowest CVaR\(_{\alpha}\) across the given scenarios.

Once the minimum-CVaR portfolio is found at iteration \(k\), the framework constructs a “synthetic asset” whose returns are precisely the weighted portfolio returns under each scenario:
\[
r_{\mathrm{AMCVaRP},k}[s] 
\;=\;
\mathbf{R}_k[s,:] \,\mathbf{w}_k^{\star},
\quad
s=1,\ldots,S.
\]
These synthetic returns are appended as an additional column to the original matrix, producing
\[
\mathbf{R}_{k+1} 
\;=\;
[\,
\mathbf{R}_k,\;
r_{\mathrm{AMCVaRP},k}
\,],
\]
thereby enlarging the asset universe to \(N_{k+1} = N_k + 1\) assets. A subsequent iteration then re-solves the minimum-CVaR optimization on \(\mathbf{R}_{k+1}\). Convergence is assessed by comparing consecutive CVaR values or by verifying whether the portfolio weights cease to change meaningfully. A typical convergence criterion might be
\[
\bigl|\,
\text{CVaR}_{\alpha}(\mathbf{w}_k^{\star})
\;-\;
\text{CVaR}_{\alpha}(\mathbf{w}_{k+1}^{\star})
\bigr|
\;<\;
\epsilon,
\]
where \(\epsilon\) is a small tolerance level indicating that further improvements in tail-risk reduction are negligible.

\section{Structural Breaks based on the AMRR derived from Cryptocurrencies}

\begin{figure}[h]
    \centering
    \includegraphics[width=0.85\linewidth]{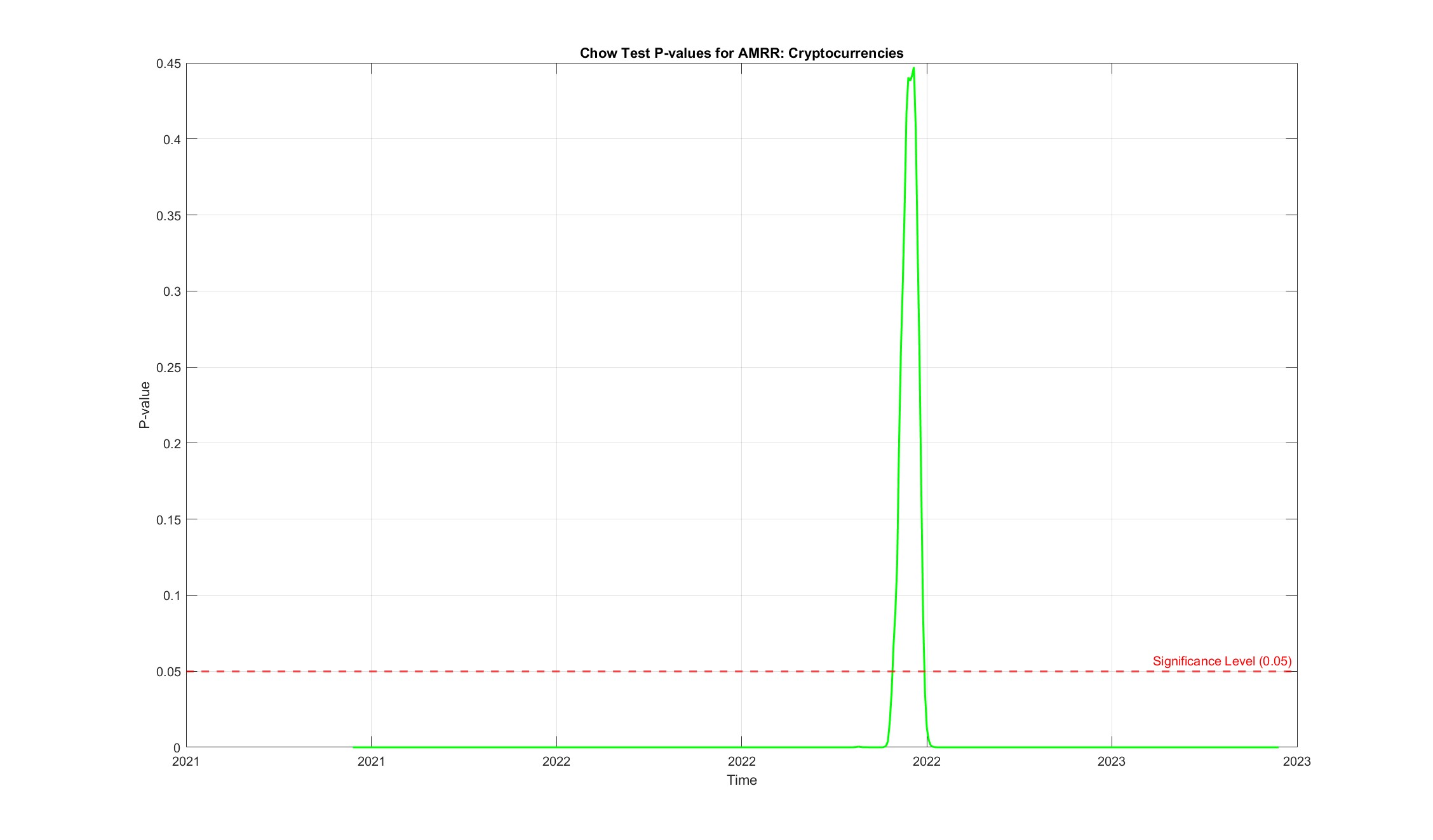}
    \caption{Chow Test Results for AMRR based on the Portfolio of Cryptocurrencies. The plot identifies a significant structural break in 2022, highlighting a critical change in the dynamics governing the measure of shadow rate. In the context of cryptocurrencies, the structural break observed in 2022 aligns with well-documented market turbulence during that period. The cryptocurrency market experienced a sharp decline in valuation, commonly referred to as the ``crypto winter," which was triggered by several interrelated factors. Notably, tightening monetary policy by central banks, including the Federal Reserve's rate hikes, reduced the liquidity available for riskier asset classes, such as cryptocurrencies. Additionally, market-specific shocks, including the collapse of high-profile stablecoins (e.g., TerraUSD) and the associated fallout from hedge funds and exchanges, exacerbated volatility and eroded investor confidence. This period marked a transition in market sentiment, shifting from speculative euphoria to risk aversion, as reflected in heightened volatility and significant price corrections across most digital assets.}
    \label{fig:chow crypto}
\end{figure}

\begin{figure}[h]
    \centering
    \includegraphics[width=1\linewidth]{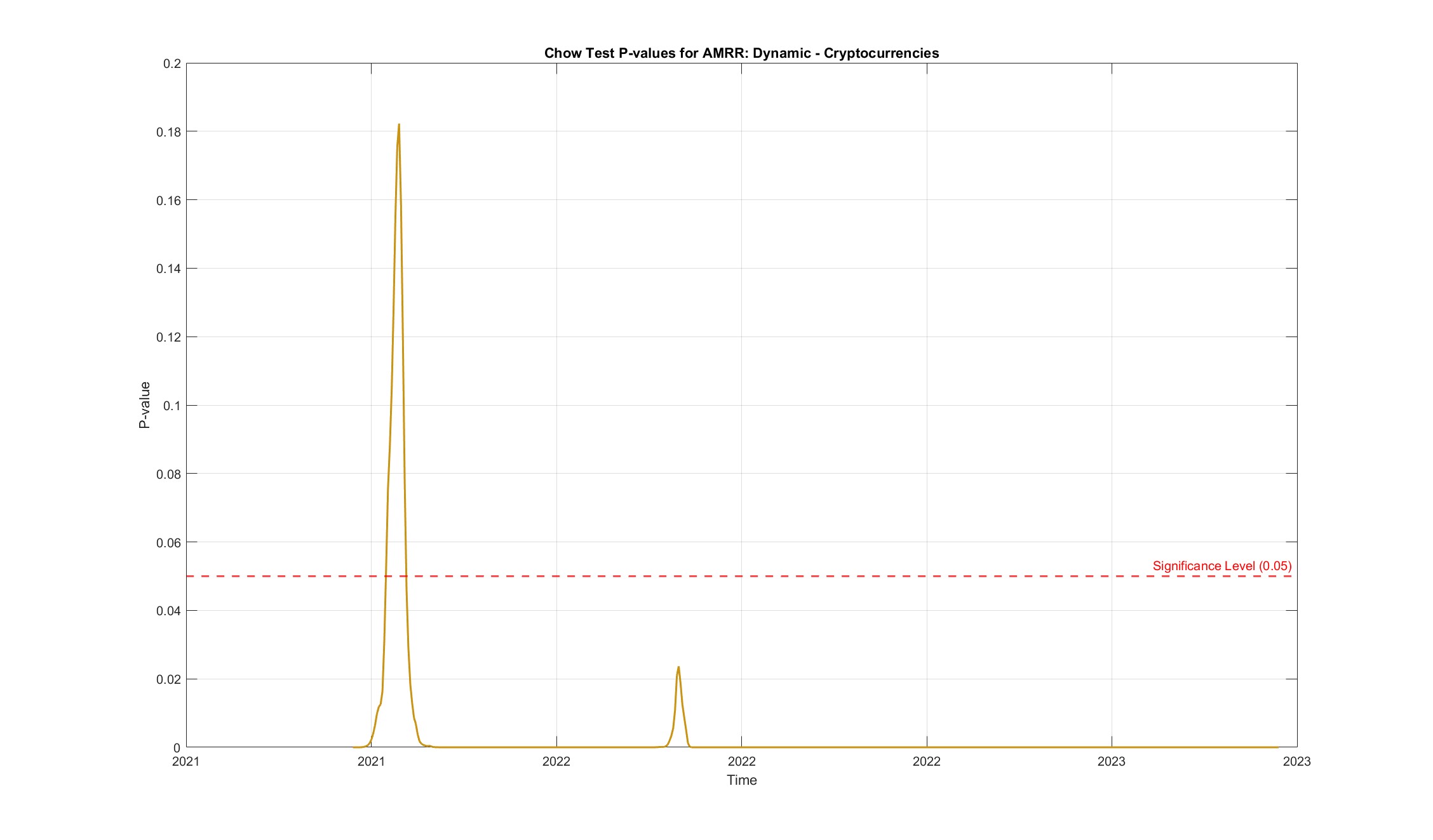}
    \caption{Chow Test Results for foward-looking AMRR based on portfolio of Cryptocurrencies.  A significant structural break is identified in 2021, marking a departure from the structural break observed in the backward-looking measure of the AMRR for the cryptocurrency portfolio. The divergence in the timing of structural breaks between the forward-looking and backward-looking AMRR measures arises from the differences in how the two approaches model asset return distributions and market dynamics. The structural break detected in 2021 aligns with a period of heightened volatility and significant market events in the cryptocurrency market. This period included the dramatic rise and subsequent correction in the value of major cryptocurrencies such as Bitcoin and Ethereum, driven by speculative trading, regulatory uncertainties, and technological developments. Events such as China's crackdown on cryptocurrency mining, announcements regarding institutional adoption, and fluctuations in liquidity contributed to sharp movements in cryptocurrency prices. These events introduced nonlinear dependencies and large shifts in asset correlations, which were effectively captured by the NIG-based forward-looking measure but may not have been as pronounced in backward-looking historical data.}
    \label{fig:chow dc}
\end{figure}

\newpage

\section{Iterative Variance Convergence based on the AMVP Alogrithm} \label{A}

\begin{figure}[htbp]
    \centering
    \includegraphics[width=1\linewidth]{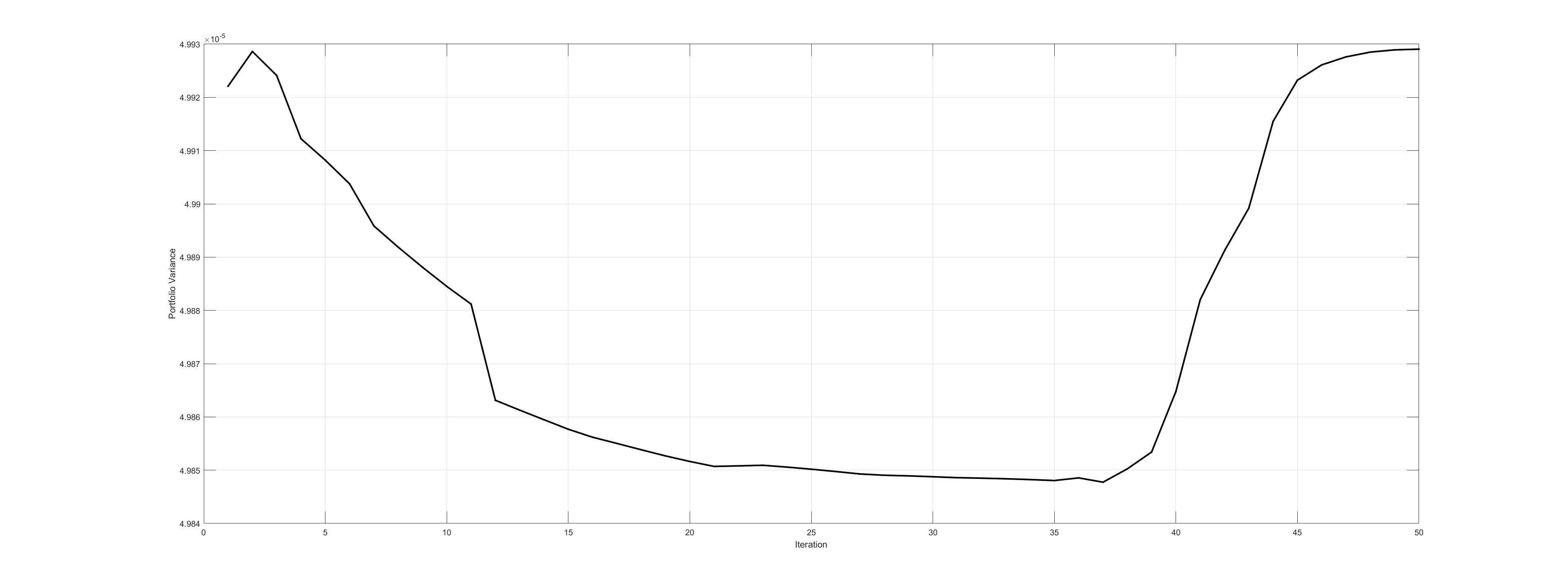}
    \caption{Iterative Variance Convergence of AMVP based on Portfolio of Assets in DJIA ($\varepsilon = 1e-20$)}
    \label{fig:convd}
\end{figure}

\begin{figure}[htbp]
    \centering
    \includegraphics[width=1\linewidth]{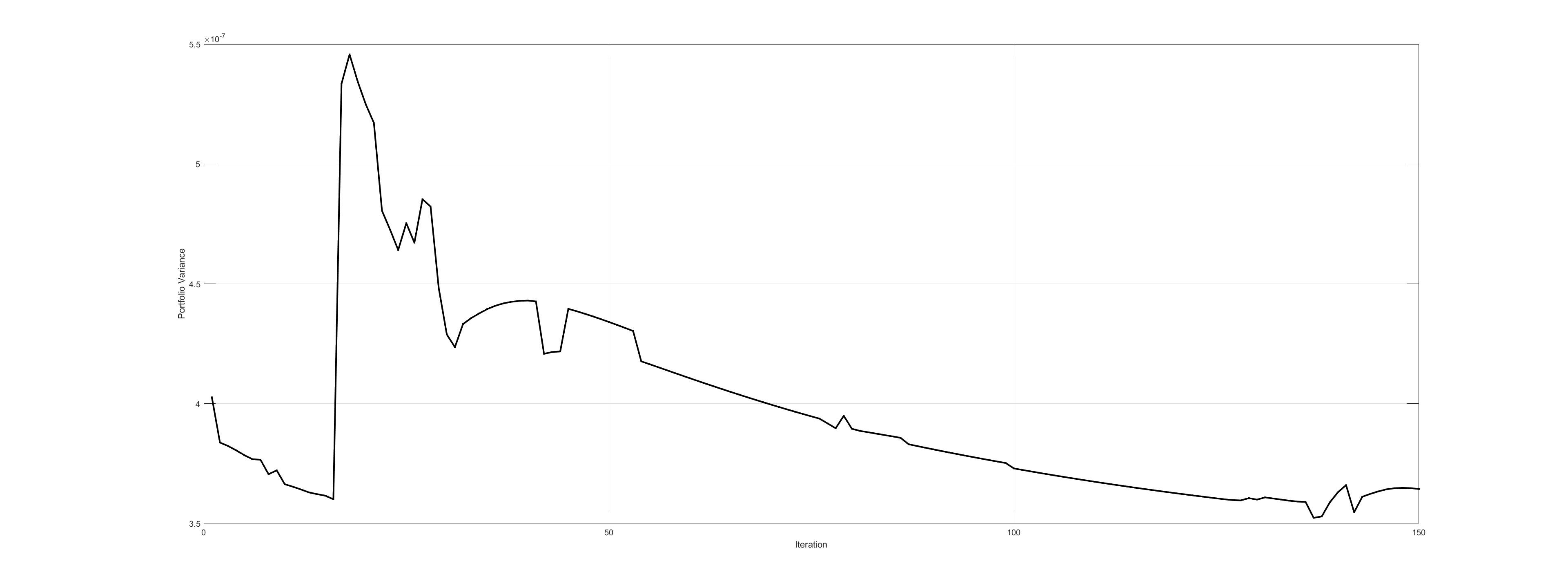}
    \caption{Iterative Variance Convergence of AMVP based on Portfolio of Cryptocurrencies ($\varepsilon = 1e-20$)}
    \label{fig:convc}
\end{figure}

\begin{figure}[htbp]
    \centering
    \includegraphics[width=1\linewidth]{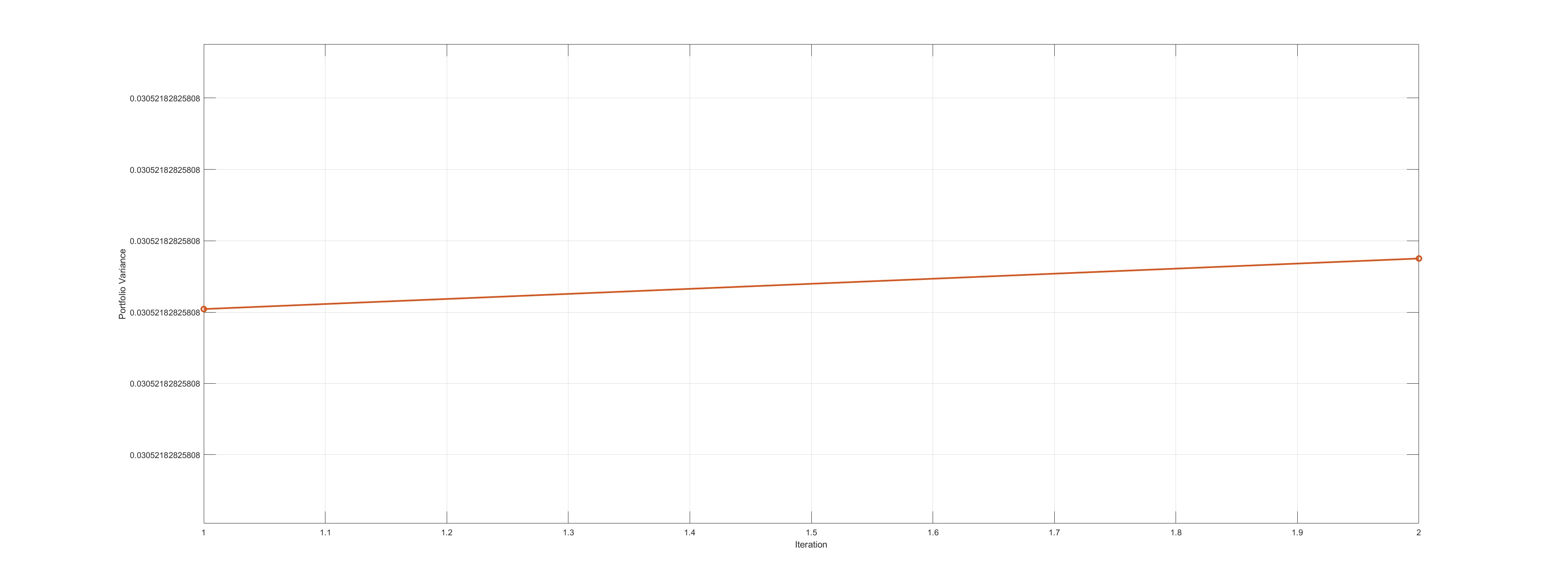}
    \caption{Iterative Variance Convergence of AMVP based on \emph{Forward-Looking} Portfolio of Assets in DJIA ($\varepsilon = 1e-6$)}
    \label{fig:convdd}
\end{figure}

\begin{figure}[htbp]
    \centering
    \includegraphics[width=1\linewidth]{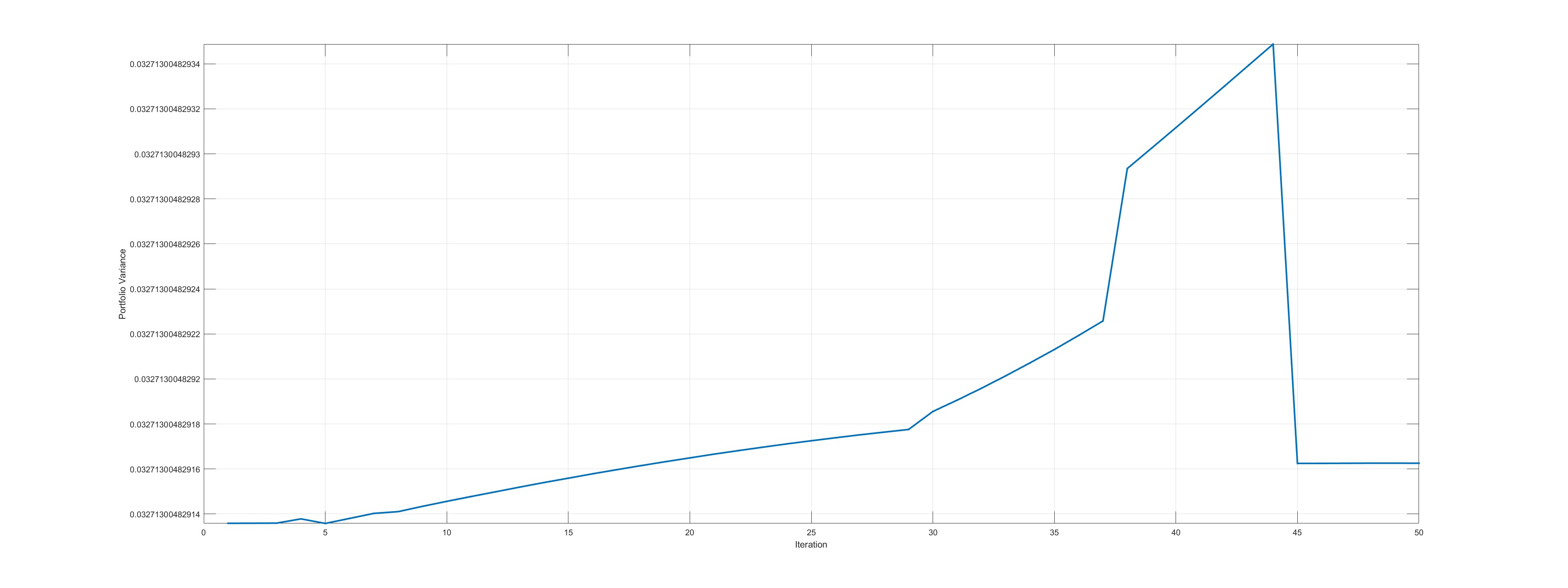}
    \caption{Iterative Variance Convergence of AMVP based on \emph{Forward-Looking} Portfolio of Cryptocurrencies ($\varepsilon = 1e-6$)}
    \label{fig:convcc}
\end{figure}

\pagebreak

\section{Summary Statistics} \label{B}

\begin{table}[h]
\centering
\begin{tabular}{lrrrrrrrr}
\toprule
Summary & MMM & AXP & AMGN & AAPL & BA & CAT & CVX & CSCO \\
\midrule
Mean & 0.000037 & 0.001090 & 0.000318 & 0.000810 & -0.000064 & 0.000959 & 0.000765 & 0.000438  \\
Std & 0.017199 & 0.018136 & 0.013903 & 0.016966 & 0.023247 & 0.018027 & 0.016485 & 0.014018  \\
Min & -0.110350 & -0.086184 & -0.072198 & -0.058680 & -0.104701 & -0.070202 & -0.067205 & -0.137304  \\
25\% & -0.008474 & -0.009235 & -0.007219 & -0.008280 & -0.012433 & -0.009064 & -0.008234 & -0.007038  \\
50\% & -0.000109 & 0.000893 & 0.000033 & 0.001095 & -0.000470 & 0.001017 & 0.001066 & 0.000410  \\
75\% & 0.007847 & 0.011177 & 0.007858 & 0.010301 & 0.012543 & 0.010603 & 0.009522 & 0.007579  \\
Max & 0.229906 & 0.105402 & 0.118180 & 0.088975 & 0.094630 & 0.088547 & 0.089035 & 0.068002  \\
\bottomrule
\end{tabular}
\caption{Assets in DJIA (Part 1)}
\end{table}

\begin{table}[h]
\centering
\begin{tabular}{lrrrrrrrr}
\toprule
Summary & KO & DOW & GS & HD & HON & IBM & INTC & JNJ \\
\midrule
Mean  & 0 .000211 & -0.000076 & 0.001099 & 0.000525 & 0.000205 & 0.000716 & -0.000270 & 0.000107  \\
Std &  0.009839 & 0.016376 & 0.016716 & 0.014903 & 0.012710 & 0.013809 & 0.025478 & 0.010102 \\
Min & -0.069626 & -0.060583 & -0.069670 & -0.088506 & -0.076174 & -0.099050 & -0.260585 & -0.039833 \\
25\% & -0.005089 & -0.009469 & -0.008816 & -0.007867 & -0.006980 & -0.006041  & -0.012913 & -0.005512 \\
50\% & 0.000781 & 0.000000 & 0.000414 & 0.000976 & 0.000802 & 0.001030 & 0.000000 & 0.000169 \\
75\% & 0.005720 & 0.009645 & 0.010665 & 0.009160 & 0.007356 & 0.007608 & 0.013174 & 0.005704 \\
Max & 0.038671 & 0.063651 & 0.130978 & 0.087010 & 0.044712 & 0.094866 & 0.106585 & 0.060728 \\
\bottomrule
\end{tabular}
\caption{Assets in DJIA (Part 2)}
\label{tab:my_label}
\end{table}

\begin{table}[h]
\centering
\begin{tabular}{lrrrrrrrrr}
\toprule
 Summary & JPM & MCD & MRK & MSFT & NKE & NVDA & PG  \\
\midrule
Mean & 0.000874 & 0.000365 & 0.000315 & 0.000814 & -0.000358 & 0.002907 & 0.000263  \\
Std  & 0.015387 & 0.010680 & 0.013089 & 0.016413 & 0.020701 & 0.033159 & 0.010612  \\
Min &  -0.064678 & -0.051225 & -0.098630 & -0.077156 & -0.199809 & -0.100046 & -0.062322  \\
25\% & -0.007651 & -0.005763 & -0.006370 & -0.007681 & -0.010039 & -0.016025 & -0.005641 \\
50\%  & 0.000957 & 0.000335 & 0.000364 & 0.000638 & -0.000185 & 0.003190 & 0.000589 \\
75\%  & 0.009212 & 0.006246 & 0.007048 & 0.010498 & 0.010042 & 0.021723 & 0.006447  \\
Max & 0.115445 & 0.040876 & 0.083744 & 0.082268 & 0.155314 & 0.243696 & 0.042699 \\
\bottomrule
\end{tabular}
\caption{Assets in DJIA (Part 3)}
\end{table}

\begin{table}[h]
\centering
\begin{tabular}{lrrrrrrrrr}
\toprule
Summary & CRM & TRV & UNH & VZ & V & WBA & WMT \\
\midrule
Mean  & 0.000476 & 0.000764 & 0.000664 & -0.000262 & 0.000524 & -0.001200 & 0.000592 \\
Std  & 0.022635 & 0.014028 & 0.014101 & 0.013103 & 0.014354 & 0.022680 & 0.012280 \\
Min & -0.197371 & -0.077607 & -0.081120 & -0.074978 & -0.069192 & -0.221584 & -0.113757 \\
25\% & -0.010878 & -0.006717 & -0.006871 & -0.006338 & -0.007163 & -0.011993 & -0.005375 \\
50\% & 0.000578 & 0.001372 & 0.000799 & 0.000000 & 0.000929 & -0.001352 & 0.000743 \\
75\% & 0.012322 & 0.008521 & 0.008150 & 0.006159 & 0.007832 & 0.010021 & 0.007150 \\
Max & 0.114969 & 0.090019 & 0.072407 & 0.092705 & 0.105991 & 0.157778 & 0.069865 \\
\bottomrule
\end{tabular}
\caption{Assets in DJIA (Part 4)}
\end{table}

\begin{table}[h]
    \centering
    \begin{tabular}{lrrrrrrrr}
\toprule
Summary & BTC- & ETH- & USDT- & BNB- & SOL- & XRP- & HBAR- & USDC- \\
& USD & USD & USD & USD & USD & USD & USD & USD \\
\midrule
Mean & 0.001649 & 0.002198 & 0.000000 & 0.003191 & 0.005287 & 0.002576 & 0.002898 & 0.000000 \\
Std & 0.032752 & 0.042047 & 0.000620 & 0.047526 & 0.063867 & 0.059968 & 0.062979 & 0.001070 \\
Min & -0.159747 & -0.272003 & -0.011296 & -0.332656 & -0.422809 & -0.423340 & -0.343541 & -0.027994  \\
25\% & -0.013768 & -0.017415 & -0.000166 & -0.015052 & -0.030125 & -0.020788 & -0.027041 & -0.000115  \\
50\% & 0.000186 & 0.001030 & -0.000006 & 0.001285 & 0.000533 & 0.000241 & 0.000243 & 0.000003 \\
75\% & 0.016551 & 0.021766 & 0.000143 & 0.018835 & 0.034984 & 0.020460 & 0.024810 & 0.000118  \\
Max & 0.187465 & 0.259475 & 0.009164 & 0.697604 & 0.357030 & 0.730750 & 0.730876 & 0.021172 \\
\bottomrule
\end{tabular}
\caption{Assets in Portfolio of Cryptocurrencies (Part 1)}
\end{table}

\begin{table}[h]
    \centering
    \begin{tabular}{lrrrrrrrr}
\toprule
Summary & OM- & ADA- & CRO- & TRX- & FET- & AVAX- & POL- & DAI- \\
& USD & USD & USD & USD & USD & USD & USD & USD \\
\midrule
Mean  & 0.005165 & 0.002591 & 0.002066 & 0.002259 & 0.004802 & 0.003623 & 0.002866 & 0.000000 \\
Std & 0.077543 & 0.050822 & 0.054718 & 0.041430 & 0.073000 & 0.065640 & 0.096707 & 0.001501 \\
Min & -0.444530 & -0.260094 & -0.277439 & -0.318372 & -0.354661 & -0.364950 & -0.364435 & -0.025125 \\
25\% & -0.027980 & -0.023449 & -0.021150 & -0.012463 & -0.036709 & -0.029718 & -0.020577 & -0.000285 \\
50\% & 0.000369 & -0.000063 & 0.000350 & 0.002300 & -0.001542 & 0.000038 & -0.000000 & -0.000009 \\
75\% & 0.030408 & 0.023070 & 0.021616 & 0.015939 & 0.038420 & 0.031667 & 0.008974 & 0.000282 \\
Max & 0.685650 & 0.322384 & 0.653658 & 0.396847 & 0.435398 & 0.750013 & 1.615706 & 0.019218 \\
\bottomrule
\end{tabular}
\caption{Assets in Portfolio of Cryptocurrencies (Part 2)}
\end{table}

\begin{table}[h]
    \centering
    \begin{tabular}{lrrrrrrr}
\toprule
Summary & WBTC-USD & XMR-USD & WETH-USD & LINK-USD & BCH-USD & DOT-USD & DOGE-USD \\
\midrule
Mean & 0.001650 & 0.001195 & 0.002179 & 0.001511 & 0.001806 & 0.001592 & 0.007143 \\
Std & 0.032739 & 0.044273 & 0.041677 & 0.053356 & 0.053950 & 0.053862 & 0.118663 \\
Min & -0.160204 & -0.413860 & -0.264302 & -0.372388 & -0.352484 & -0.379335 & -0.402570 \\
25\% & -0.013057 & -0.016608 & -0.017859 & -0.027777 & -0.023029 & -0.026533 & -0.025541 \\
50\% & 0.000602 & 0.002481 & 0.001024 & 0.001407 & 0.000068 & -0.000361 & -0.000379  \\
75\% & 0.016526 & 0.020775 & 0.021902 & 0.029974 & 0.022080 & 0.025210 & 0.023912 \\
Max  & 0.193579 & 0.411925 & 0.263107 & 0.317596 & 0.584202 & 0.376014 & 3.555466 \\
\bottomrule
\end{tabular}
\caption{Assets in Portfolio of Cryptocurrencies (Part 3)}
\end{table}

\begin{table}[h]
    \centering
    \begin{tabular}{lrrrrrrr}
\toprule
Summary & LEO-USD & NEAR-USD & XLM-USD & LTC-USD & ETC-USD & BTCB-USD & UNI7083-USD \\
\midrule
Mean & 0.001777 & 0.003552 & 0.002117 & 0.001144 & 0.002628 & 0.001780 & 0.002327 \\
Std & 0.033696 & 0.067116 & 0.055848 & 0.046311 & 0.057567 & 0.037045 & 0.059613 \\
Min & -0.180933 & -0.358289 & -0.303946 & -0.356729 & -0.322014 & -0.189450 & -0.332005 \\
25\% & -0.013057 & -0.016608 & -0.017859 & -0.027777 & -0.023029 & -0.026533 & -0.025541 \\
50\% & 0.000667 & -0.000543 & 0.000584 & 0.001382 & 0.000392 & 0.000643 & -0.000067 \\
75\% & 0.010281 & 0.036144 & 0.019683 & 0.022580 & 0.022849 & 0.016090 & 0.028268 \\
Max & 0.554760 & 0.434801 & 0.749244 & 0.282016 & 0.422583 & 0.371626 & 0.543089 \\
\bottomrule
\end{tabular}
\caption{Assets in Portfolio of Cryptocurrencies (Part 4)}
\end{table}

\pagebreak

\begin{sidewaysfigure}[h]
    \centering
    \includegraphics[width=0.9\linewidth]{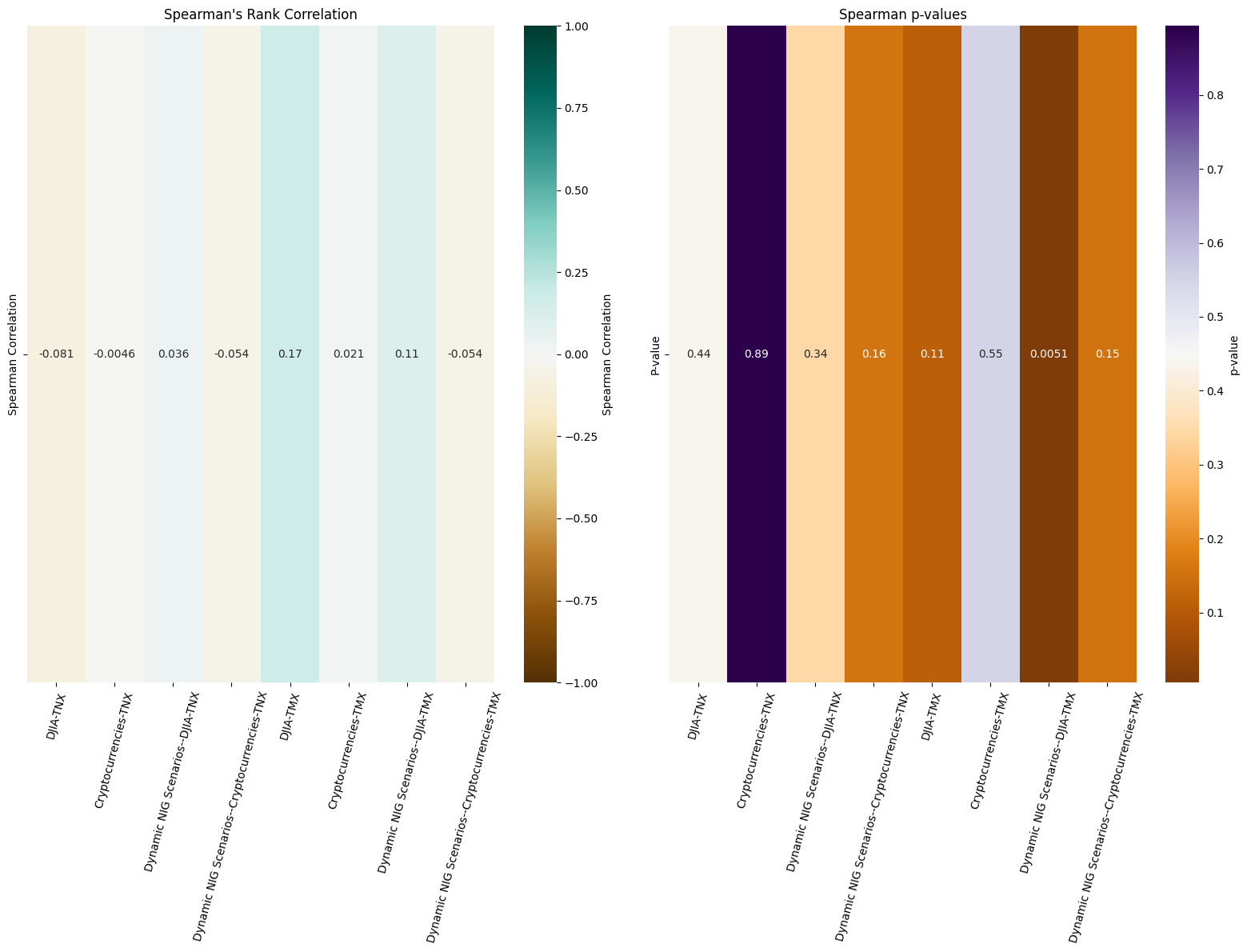}
    \caption{Spearman's Rank Correlation \\ Figure \ref{fig:correlation} display the Spearman’s rank correlation coefficients (left) and associated p-values (right) for four log-transformed AMRR series—two derived from DJIA portfolios (static vs. dynamic NIG scenarios) and two from cryptocurrency portfolios (static vs. dynamic NIG scenarios)—each measured against two Treasury rates: 10Y (TNX) and 3M (TMX). The small magnitudes of the correlations (mostly between \(-0.08\) and \(+0.17\)) and high p-values suggest there is no strong co-movement in most cases. An exception emerges for the dynamic DJIA AMRR correlated with the 3M rate, which shows a correlation of approximately 0.11 and a p-value near 0.005, indicating a slight but statistically significant positive relationship. This implies a marginal sensitivity of the dynamic DJIA AMRR to short-term interest rates, whereas no similar evidence is observed for the 10Y rate or for the AMRR measures derived from cryptocurrency portfolios.}
    \label{fig:correlation}
\end{sidewaysfigure}

\end{document}